\documentclass[amsmath,amssymb,floatfix,prb,epsf,eqsecnum,showpacs,twocolumn]{revtex4}
\usepackage{amsmath,amssymb,natbib,bm,graphicx,url,epsfig}

\newcommand{\be}{\begin{equation}}
\newcommand{\ee}{\end{equation}}
\newcommand{\bes}{\begin{equation}\begin{split}}
\newcommand{\ees}{\end{split}\end{equation}}
\newcommand{\ba}{\begin{eqnarray}}
\newcommand{\ea}{\end{eqnarray}}

 \DeclareMathOperator{\sgn}{sgn}

\def\beq{\begin{equation}}
\def\eeq{\end{equation}}
\def\bea{\begin{eqnarray}}
\def\eea{\end{eqnarray}}

\begin{document}
\title{Fermionic propagators for 2D systems with singular interactions}
\author{Tigran A. Sedrakyan and Andrey  V.~Chubukov}
 \affiliation{Department of Physics, University of Wisconsin-Madison, Madison,
Wisconsin 53706, USA}


\begin{abstract}
We analyze the form of the fermionic propagator for
 2D fermions interacting with massless overdamped bosons.
 Examples include a nematic and Ising
 ferromagnetic quantum-critical points, and fermions at a
 half-filled Landau level. Fermi liquid behavior in these systems
 is broken at criticality by a singular self-energy, but
 the Fermi surface remains well defined. These are strong-coupling
 problems with no expansion parameter other than
 the number of fermionic species, $N$. The two known limits, $N >>1$ and
$N=0$ show qualitatively different behavior of the fermionic propagator
 $G(\epsilon_k, \omega)$. In
 the first limit, $G(\epsilon_k, \omega)$ has a pole at some
$\epsilon_k$, in the other it is analytic. We analyze the crossover
between the two limits.  We show that the pole survives for all $N$,
but  at small $N$
it only exists in a range
$O(N^2)$ near the mass shell. At larger distances from the mass shell, the system evolves and $G(\epsilon_k, \omega)$ becomes regular. At $N=0$, the range
 where the pole exists collapses and $G(\epsilon_k,
\omega)$ becomes regular everywhere.

\end{abstract}

\maketitle
\section{Introduction}

Physical properties of
fermionic systems interacting with critical neutral fluctuations
have been a focus of intense studies over the last several decades
and yet remain a subject of advanced research.
Examples include fermions interacting with a gauge field,~\cite{Lee,AIM}
 a half-filled Landau level,~\cite{AIM,senthil}
and the system behavior at  quantum-critical points (QCP)
 towards Ising-type ferromagnetism~\cite{Chubukov04,Chubukov06,Dzero} (FM) and
towards a nematic order.\cite{Oganesyan,Lawler,MRA,Kee}
The later case is an example of a
 Pomeranchuk-type Fermi surface
 instability of an isotropic  Fermi liquid\cite{Pom1}.
In all such systems, scattering of fermions by
massless bosonic excitations leads to a non-analytic form of the
 fermionic self-energy $\Sigma (k, \omega_m)$. Below the upper critical
dimension $D_{cr}$, $\Sigma (k_{\text{\tiny{F}}}, \omega_m)$ exceeds a bare $i\omega_m$  term in the fermionic propagator, and the system develops
 a non-Fermi liquid behavior. At a critical point towards a nematic or an Ising ferromagnet,  one-loop self-energy $\Sigma (k_{\text{\tiny{F}}}, \omega_m) \propto i \omega_m^{D/3}$, and $D_{cr} =3$ (Ref. \onlinecite{Lee}).  The
  singular behavior of the self-energy is, however,
 only in the frequency domain, the momentum dependence of $\Sigma (k, 0)$
remains regular: $\Sigma (k, 0) \propto \epsilon_k$.
 As a consequence, the Fermi surface remains well defined at $k_{\text{\tiny{F}}}$
as a locus of singular points of $G({\bf k}, 0)$ despite that Landau quasiparticles do not exist.

In 2D, one-loop self-energy $\omega^D_m$ becomes $\omega^{2/3}_m$.
 It has long been the issue~\cite{AIM,Khvesh,Chubukov06,Lawler,CK06}
whether $\omega^{2/3}_m$
form is the exact expression for a non-Fermi liquid fermionic propagator.
The answer to this question is still lacking. On one hand, the
two-loop and higher order self-energies also scale as $\omega_m^{2/3}$, i.e., the exponent remains the same to all orders.
On the other hand, higher-order terms are of the same order as one-loop self-energy, and it is a'priori unclear what the sum of infinite series of $\omega_m^{2/3}$ terms yields.

The way to treat such systems in a controlled way is to artificially
 extend them to $N$ fermionic flavors and require that the interaction with a boson conserves the flavor. At large $N$, multi-loop $\omega_m^{2/3}$ self-energy terms acquire extra powers of $(\ln N/N)^2$ and the series of $\omega_m^{2/3}$ terms converge. In this situation,  self-energy is essentially determined by the one-loop term $\Sigma (k, \omega_m) = i |\omega_m|^{2/3} \omega^{1/3}_0
 \sgn (\omega_m)$, where
$\omega_0$ is the internal energy scale. Accordingly, at $\omega_m>0$ which we only consider below,
\bea
G({\bf k}, \omega_m) &\approx& \frac{1}{i \omega^{2/3}_m \omega^{1/3}_0 - \epsilon_k} \nonumber\\
&=&
-\frac{1}{\epsilon_k} F_{N \to \infty}  \left(-\frac{\omega^{2/3}_m \omega^{1/3}_0}{\epsilon_k} \right), \nonumber \\
 F_{N \to \infty} (x) &= & \frac{1}{1+ix}.
\label{n_1}
\eea
As a function of a complex $x$, $F_{N\rightarrow\infty}(x)$ has a simple pole at $x = i$.
Because of the pole, real space/time propagator $G({\bf r},t)$   is long-ranged and decays by a power-law, as  $G({\bf r},t=0) \propto 1/r^2$
and $G(v_F t >> {\bf r})\propto 1/(t\sqrt{r})$.
 (see Appendix~\ref{pole}).

Another solvable limit is $N=0$. In this case
 the curvature of the 2D Fermi surface scales out,
 the system behavior becomes effectively one-dimensional
 and can be obtained by bosonization.  The Green's function at $N=0$ has been obtained by Ioffe {\em et al}~~\cite{Lidsky} and Altshuler {\em et al}.~\cite{AIM}
It is still given by Eq.~(\ref{n_1}), but the functional form of $F_{N=0} (x)$
is fundamentally different: $F_{N=0} (x)$ is analytic in any
 finite region in the upper half-plane of $x$ and becomes singular only at $x = \infty$. Because
 the pole is absent, $G({\bf r},t)$ is now short-ranged and
 at $v_{\text{\tiny{F}}}t\gg r$ scales as $G({\bf r},t)
 \propto \left[1/t\sqrt{r}\right]\exp [-a r/t^{2/3}]$, where $a$ is a dimensional prefactor.

Different behavior of $F_N (x)$ at large $N$ and at $N=0$ raises the question which of the two forms (if any) describes  system behavior in the physical case of $N=1$.   Altshuler {\em et al} conjectured~\cite{AIM} that
the $N=0$ case is special, and
 the system behavior at any $N \neq 0$ is qualitatively the same as for large $N$, i.e.,  the pole in $F_N$ persists for all $N$
 and only vanishes
at
 $N=0$.  On the contrary,
Fradkin and Lawler~\cite{Lawler} argued that exponential behavior of $G({\bf r},t)$
 survives at finite $N$. One of us and Khveshchenko argued~\cite{CK06}
that the curvature of a 2D Fermi surface is relevant for any $N >0$, and $G({\bf r},t)$ decays by a power-law once $N>0$. However, the calculations in Ref.~\onlinecite{CK06}  are approximate and did not yield the same $G({\bf r},t=0)\propto 1/r^2$
as at $N \rightarrow \infty$.

In this paper, we analyze this issue in detail by performing loop expansion at small $N$. We find, in agreement with the conjecture by Altshuler {\em et al}~~\cite{AIM} that the pole in the Green's function exists at any finite $N$, and real-space, equal time Green's function $G(r, t=0)$ decays by a power-law, as $1/r^2$. The way how the pole disappears at $N \rightarrow 0$ is, however,
 somewhat counter-intuitive. Naively, one could expect that the residue of the pole $Z_N$ gradually vanishes as $N\rightarrow 0$. We, however, found a different behavior: the residue $Z_N$ remains $O(1)$ at small $N$, but the
 pole only exists in the range $\Delta < N^2$, where $\Delta$, introduced in (\ref{y_1}) below, is the
 distance from the pole. Outside this range, the Green's function is regular and the same as at $N=0$. At $N \rightarrow 0$, the range collapses and the Green's function becomes regular even at $\Delta =0$.

We present computational details below, but first summarize our rational.
As our primary goal is to study what happens at small but finite $N$, we cannot use bosonization, which is only applicable at $N=0$, and have to rely on
 the diagrammatic loop expansion. It is not guaranteed a'priori that
loop expansion is useful at small $N$ as all terms in the series are
 generally of the same order, and  it could be the case that the corrections are all regular near the mass shell, but the prefactors are
 arranged such that infinite series of regular $O(1)$
corrections to the quasiparticle residue diverge at $N=0$ and
 destroy the pole.

We, however, found that the actual situation is different, and the
 the pole disappears at $N=0$ and is still present at a finite $N$
not because regular series diverge (at $N=0$) or almost diverge (at $N >0$),
but because of a peculiar singularity in the self-energy, whose form is different at $N=0$ and at  finite $N$.  This singularity
 can be captured within the loop expansion. Specifically, we find that
 the expansion of the self-energy  $\Sigma$ near the mass shell,
  in powers of
\be
\Delta = 1 - \frac{i}{x} = 1 + i \frac{\epsilon_k}{\omega^{2/3}_m \omega^{1/3}_0}\label{y_1}
\ee
 contains a
universal term  which is ``non-perturbative'' in the sense that it  comes
 from fermions whose energies are of order $\omega \Delta^{3/2}$, which are
  smaller than the external energy $\omega$.
This term appears at the two-loop order and at $N=0$ yields the non-analytic
contribution to the   self-energy  which scales as $\Delta^{3/2}$. At a first glance, such
 term cannot eliminate a pole as it is smaller than
$G^{-1}_0 = \Delta$. However, this universal self-energy gets renormalized by
 logarithmically divergent vertex corrections, which, because typical energies scale with $\Delta$, are powers of $\ln\Delta$.
  Series of such corrections
exponentiate into the full vertex $\Gamma \propto \Delta^{-a}$ and
  modify the universal self-energy to  $\Delta^{3/2} \Gamma^2 \sim \Delta^{3/2-2a}$. The exponent $a=1$ in the leading logarithmical  approximation (when only
the highest power of the logarithm is kept at any order), but gets renormalized by $O(1)$ corrections beyond this approximation.
We cannot find $a$ explicitly, but a comparison with bosonization implies that
 $a=3/4$ in which case the fully renormalized self-energy tends to a constant value on the mass shell, i.e., the pole in $G(k, \omega)$ disappears.

We next turn to $N >0$. We found that universal contribution to the
 self-energy do exist in this case as well, but at the smallest $\Delta << N^2$
 it behaves as $\Delta^{5}/N^2 $
 (up to extra logarithms)
 Because typical internal energies are still
 small, vertex corrections are again relevant, but now  $\ln \Delta$
 gets replaced by $\ln N^2$ for $\Delta << N^2$, such that
$\Gamma \propto 1/N^{3/2}$. The full universal self-energy is then $(\Delta^{5/2}/N^2) \Gamma^2 \sim (\Delta/N^2)^{5/2}$. As the result, the pole survives, at the smallest $\Delta$, and  its  residue remains $O(1)$. However,
$(\Delta/N^2)^{5/2}$ form of the self-energy is only valid for $\Delta/N^2$. At larger deviations from mass shell,
the self-energy approaches the same constant
value as at $N=0$. Generally, at small $N$, we have with logarithmic accuracy,
\bea
&&G(k, \omega_m) = -\frac{1}{\epsilon_k} F_{N \rightarrow 0}
\left(-\frac{\omega^{2/3}_m \omega^{1/3}_0}{\epsilon_k}\right) \nonumber \\
&&  F_{N \rightarrow 0} (x) =  F_{N \rightarrow 0} \left (\Delta\right) = \frac{1}{-\Delta + g\left(\Delta/N^2\right)},
\label{y_5}
\eea
where
$ g(y) \propto y^{5/2}$ at  $y \ll 1$, and  $g (y) \approx {\text const}$ at
$ y \gg 1$.

To re-iterate, our key point is that there exist a universal,
 non-analytic term in the loop expansion for the self-energy.
At a finite $N$, this term is cast into
 the scaling function of $\Delta/N^2$, where
$\Delta$ is the deviation from the mass shell. At the smallest $\Delta$, it
 scales as $\Delta^{5/2} \ll \Delta$, and the
 pole in $G(k, \omega_m)$ then survives, with the residue $Z+N = O(1)$.
At $\Delta > N^2$ the self-energy approaches a constant, and
 the full $G(k, \omega_m)$ looses the memory about the pole.
At $N=0$, the region where the pole exists vanishes, and the Green's function becomes regular even at $\Delta = 0$.

We also considered another
 example of a non-Fermi liquid behavior - the  case of 2D
fermions at the the half-filled $(\nu=1/2)$ Landau level.~\cite{Jain,Halperin,AIM} In this case, the fermionic self-energy is marginal  at large $N$,
 $\Sigma (k, \omega_m) = i \lambda \omega_m  |\ln \omega_m|$, where $\lambda$ is a dimensionless coupling (see below). The fermionic
 propagator $G({\bf k}, \omega_m) = -(1/\epsilon_k) \tilde{F}_N (-y)$, with
 $y = \lambda \omega_m  |\ln \omega_m|/\epsilon_k$ then
 again has a pole at $y=i$.
We solved the $N=0$ limit by bosonization and found that
$ \tilde{F}_0 (y)=e^{-iy}$,
which is obviously a regular function along imaginary $y$ axis.
 We then  performed  the small $N$ analysis
 and found the behavior which is similar but not equivalent to the previous case. Namely, at the distance from the pole ${\tilde \Delta} =
1 + i \epsilon_k/(\lambda \omega_m |\ln \omega_m|) \ll N^2$,
the pole still exists at any finite $N$, but its residue now scales as
 $Z_N \propto N^2$.
At larger deviations from the mass shell, the universal self-energy approaches
a constant, and the system looses the memory about the pole. At $N=0$, the range where the pole exists collapses, and the Green's function becomes regular for all ${\tilde \Delta}$.

The paper is organized as follows. In Section~II we introduce the model. In Section~III we consider a 2D system at a Pomeranchuk QCP towards a
 nematic order. We present the results for large $N$ and briefly review
 bosonization results for $N=0$. We then discuss universal terms
 in the loop expansion at $N=0$ and at a finite $N$.
 In Section~IV we present the same consideration for
 2D electrons at the half-filled Landau level.
In Section~V we present the Conclusions.
Some technical details are presented in the Appendices.

\section{The model}

We consider 2D fermions with a circular Fermi surface and
 dispersion $\epsilon_k$.
 We assume that fermions interact at low energies
 by exchanging collective excitations with the static propagator $\chi (q) = \chi_0/q^{1+x}$.  For nematic and Ising ferromagnetic QCP  $x=1$,~\cite{AIM,Chubukov04,Chubukov06,Dzero,Oganesyan,MRA,Kee}
 for the half-filled Landau level with unscreened Coulomb interaction, $x=0$ (Refs.
 \onlinecite{Jain,Halperin,AIM}). We will only consider interactions with charge fluctuations. Interactions with gapless spin fluctuations require separate treatment.\cite{Chubukov04,Chubukov06}
 The
 static $\chi (q)$ is predominantly created by high-energy fermions and is an input for the low-energy model.\cite{acs} The Hamiltonian of the model is given by
\begin{eqnarray} \label{Act1}
{\mathcal H}&=&
\sum_{k,\alpha} \epsilon_k  c^\dagger_{k, \alpha} c_{k,
\alpha} + \sum_{q} \chi^{-1}_0(q){\Phi}_{q} {\Phi}_{-q }\nonumber\\
&+& g~ \sum_{k, q} c^\dagger_{k,\alpha} P_{\alpha \beta }
c_{k + q, \beta}\;{\Phi}_{-q},\qquad\qquad
\end{eqnarray}
where the first term is the kinetic energy of fermions, the second term is the
 potential energy of collective excitations described by $\Phi$, and the third term describes the interaction between fermions and collective modes. The coupling $g$ generally depends on momentum but can be approximated by a constant at small momenta. $P_{\alpha \beta} = \sigma^z_{\alpha\beta}$ for
 an Ising ferromagnet and $\delta_{\alpha \beta}$ for a nematic transition.
The results are equivalent in both cases, and we will only consider a nematic transition.

Near a particular ${\bf k}_{\text{\tiny{F}}}$ point, which for definiteness we direct along $x$,
\beq
\epsilon_k = v_{\text{\tiny{F}}} \left[(k_x -k_{\text{\tiny{F}}}) + k^2_y/(2k_{\text{\tiny{F}}})\right].
\label{n_4}
\eeq
The second term is due to the curvature of the 2D Fermi surface.

The interaction $g$
 appears in the perturbation theory only in even powers, in
 a combination ${\bar g} = g^2 \chi_0$.
 The dimension of $g$ is inverse mass $1/m$, the dimension of $\chi_0$ is
 $m k^2$, hence the dimension of ${\bar g}$ is energy. The model
 has a natural dimensionless
 parameter $\lambda = {\bar g}/E_{\text{\tiny{F}}}$, where $E_{\text{\tiny{F}}}$ is the Fermi energy which we assume to be of the same order as the fermionic bandwidth.
We assume, like in earlier works,~\cite{AIM,MRA,Chubukov06} that
$\lambda <<1$. This condition implies
that interaction does not take the system out of the low-energy domain, i.e.,
 low-energy behavior is well separated from the system behavior
 at energies compared to the bandwidth.

As it is customary for the problems in which fermions interact with their own collective modes, collective excitations become Landau overdamped due to  interaction with fermions, and the full dynamic susceptibility
 of the $\Phi$ field becomes
\begin{eqnarray}
\label{propag}
\chi({q},\omega)=\frac{\chi_0}{ \gamma\vert\omega/q\vert +
q^{1+x}},
\end{eqnarray}
where $\gamma = {\bar g} k_{\text{\tiny{F}}}/(\pi v^2_{\text{\tiny{F}}})$
(see Refs.~\onlinecite{Chubukov04,Chubukov06}).  The Landau damping can be included into the theory already at the bare level, all one has to do is to change from a Hamiltonian description to a description
in terms of an effective action.\cite{acs}

The model can  be extended to $N \neq 1$ fermionic flavors by adding a flavor index to fermions and keeping flavor index intact in the interaction with the $\Phi$ field.  This extension allows one to consider large $N$ and small $N$.
 At large $N$, fermionic damping
$\gamma$ scales as $N$ and is large.  Collective excitations then become
 slow modes, and their effect on fermions becomes small in $1/N$ by Migdal theorem.
This is the limit where a direct perturbative treatment is applicable.
In the opposite limit of small $N$, the momenta which mostly contribute to the fermionic
self-energy are of order $N$, such that the $k^2_y$ term in Eq.~(\ref{n_4})
 is small by $N$, and the curvature of the 2D Fermi surface becomes a small perturbation.~\cite{AIM, Lawler}
 Without the $k^2_y$ term, the
fermionic dispersion becomes purely one-dimensional, and the self-energy can be found by a bosonization technique. As we said, our key goal is to analyze the crossover between solvable large $N$ and $N=0$ limits with the aim to understand system behavior for the physical case of $N=1$.

Below we consider separately the case of short-range (screened) interaction
 for which $\chi (q) \propto 1/q^2$ (this is the case of fermions interacting with quantum-critical collective excitations), and the case of unscreened long-range interaction (electrons at the half-filled Landau level) for which $\chi (q) \propto 1/|q|$.

\begin{figure}[t]
\centerline{\includegraphics[width=35mm,angle=0,clip]{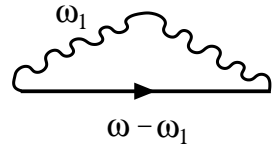}}
\caption{The one-loop
self-energy diagram.}
\label{fig:se1}
\end{figure}

\section{Short range interaction, a quantum-critical point}

The limiting cases $N \rightarrow \infty$ and $N=0$ have been studied before.
We briefly review the existing results and go one step further in the $1/N$
 expansion for large $N$. We then present and discuss our results for the universal terms in the loop expansion near the mass shell.

\subsection{Fermionic propagator at $N\gg 1$}

At large $N$, the results for the fermionic propagator can be obtained by expanding in the number of loops. Each extra order brings extra smallness in $1/N$.  Explicit calculations show that the parameter for the loop expansion
 is actually $\ln^2N/N^2$, which is even smaller than $1/N$.

The one-loop self-energy diagram is shown in Fig. \ref{fig:se1}.
The $k$ dependence of the one-loop self-energy is regular, with
 a  prefactor $O(\lambda) <<1$, and we neglect it.
The frequency dependence of $\Sigma_1 (k, \omega_m)$ is $\omega_m^{2/3}$:
\beq
\Sigma_1 (k, \omega_m) = i \omega^{2/3}_m \omega^{1/3}_0,
~~\omega_0 =  \frac{\bar{g}^2}{(2\sqrt{3})^3 \pi^2 N m v^2_{\text{\tiny{F}}}}.
\label{n_5}
\eeq
The prefactor $\omega_0$ formally contains $1/N$, but it can be absorbed into the renormalization of the Fermi velocity. All higher-order diagrams contain the same combination $\omega^{2/3}_m \omega^{1/3}_0$, and we will just consider $\omega_0$ as a normaization factor for frequency.

 The $\omega^{2/3}$ dependence of $\Sigma$ implies that quasiparticles are not sharply defined (along real frequency axis, $Re \Sigma$ and $Im \Sigma$ are of the same order $\omega^{2/3}$), i.e., the system behavior is not a Fermi liquid. At the same time, $\Sigma (k, \omega_m)$ still vanishes at zero frequency, i.e.,
the Fermi surface remains well defined.
Substituting the one-loop self-energy into $G({\bf k}, \omega_m)$ and neglecting $\omega_m$ in comparison with $\Sigma_1$ we reproduce (\ref{n_1}).

\begin{figure}[t]
\centerline{\includegraphics[width=70mm,angle=0,clip]{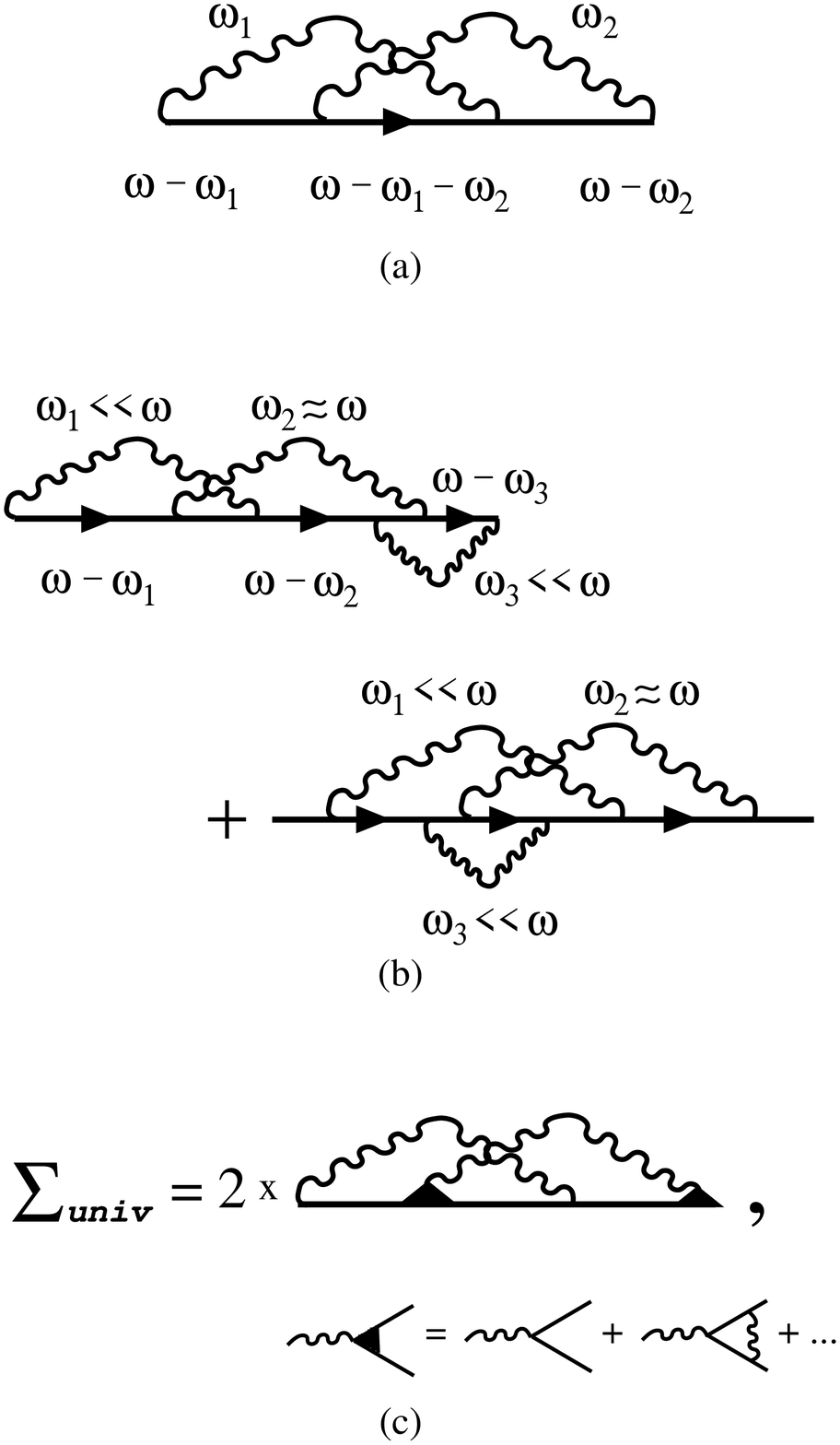}}
\caption{(a) The relevant two-loop self-energy diagrams. The universal
 contributions  come from $\omega_1, \omega -\omega_2 \sim \omega \Delta^{3/2} \ll \omega$ and
$\omega_2, \omega - \omega_1 \sim \Delta^{3/2} \omega \ll \omega$.
 (b)
Relevant vertex corrections to the two loop diagram whose universal
 contribution comes from
$\omega_1, \omega -\omega_2 \sim \omega \Delta^{3/2}$.
 For the second universal contribution, from $\omega_2, \omega - \omega_1 \sim \Delta^{3/2}$,  the corrections to the other two vertices are relevant.
 (c) The full universal self-energy with fully
renormalized vertices.
}
\label{fig:se2}
\end{figure}

The relevant two-loop self-energy diagram
is shown in Fig.~\ref{fig:se2}a. Earlier estimates show that, when $\epsilon_k =0$, $\Sigma_2 (k_{\text{\tiny{F}}}, \omega_m) \propto i \omega^{2/3}_m \omega^{1/3}_0 \bigl(\ln N/N \bigr)^2$ (Refs.~\onlinecite{AIM,Chubukov06}). We found that  $\Sigma_2 (k, \omega_m)$ is also a non-trivial
 function of $\omega^{2/3}_m \omega^{1/3}_0/\epsilon_k$.
For our purposes, we need the expression for $\Sigma_2 (\omega_m, k)$ near the
 mass shell $\Delta = 0$, where
$\Delta =  1 + i \epsilon_k/(\omega_m^{2/3} \omega_0^{1/3})$. We found (see Appendix B) that, to order $\Delta^2$, this expansion is regular and
\begin{eqnarray}
\label{quadratic2}
\!\Sigma_2 (k, \omega_m)&=&  i \omega_m^{2/3} \omega_0^{1/3}\left(\frac{\ln N}{4\pi N}\right)^2\\
&\times&\left[0.52 + \frac{\pi^2}{6}\; \Delta + O\bigl(\Delta^2\bigr)\right]\!.\nonumber
\end{eqnarray}

Adding $\Sigma_2$ to $\Sigma_1$, substituting into the Green's function
and casting $G$ into the form of Eq.~(\ref{n_1}), we find
\beq
F_N (x) = \frac{Z_N}{D_N + ix} + F_{inc} (x),
\label{n_3}
\eeq
where  $F_{inc}$ is a regular function near the mass shell $x = -i D_N$, and
\beq
Z_N = 1-2.16\left(\frac{\ln N}{4\pi N }\right)^2, ~~D_N = 1 - 0.52 \left(\frac{\ln N}{4\pi N}\right)^2.
\label{n_7}
\eeq
We see that, at this level of consideration,
 $1/N$ corrections lead to three effects: the pole in $G(\epsilon_k, x)$ acquires the residue $Z_N <1$, the location of the pole along the imaginary $x$ axis shifts to a somewhat different $x = iD_N$
 and the Green's function acquires an incoherent part which is regular near the pole.  Perturbation theory in $1/N$ is perfectly well defined, and the pole in $F_N (x)$  is surely present at large $N$.

\begin{figure}[t]
\centerline{\includegraphics[width=80mm,angle=0,clip]{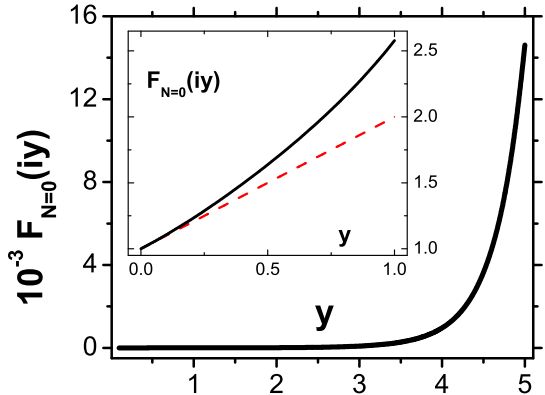}}
\caption{The function $F_{N=0}(iy)$.
Dashed line represents small $y$ asymptotics (see text).}
\label{fig:F0}
\end{figure}
\subsection{Fermionic propagator at $N=0$}

In the limit $N=0$, the curvature of the Fermi surface scales out
 and the original 2D problem maps onto an effective 1D
 theory of electrons with retarded
interaction (Refs.~\onlinecite{Lidsky,AIM}).
 This  allows one to compute fermion propagator by using
 1D bosonization technique.
The result for $G({\bf k}, \omega_m)$ can again be cast into the form of Eq.~(\ref{n_1}), but now~\cite{Lidsky,AIM}
\bea
&&F_{N=0} (x) =\frac{3}{2}\exp\left\{(-1)^{5/4}\Gamma^{3/2} (5/3)
x^{3/2}\right\}\quad\qquad\qquad\nonumber\\
&&\qquad\qquad-\frac{3\sqrt{3}i}{4\pi}\int_0^{\infty}dy\;
\frac{\exp\left\{i(\Gamma (5/3) xy)^{3/2}\right\}}{y^2+iy-1}.
\label{scaling2}
\end{eqnarray}
The analysis of $F_{N=0} (x)$ is presented in Appendix A. The key point is that $F_{N=0} (x)$ is a regular function in any finite region in the upper half-plane, and, in particular, along imaginary $x = i y, y>0$, where there was a pole at large $N$. At small $x$, $F_{N=0} (x)$ expands in regular powers of $x$:
\beq
\label{gseries1}
F_{N=0}(x=iy)=1 + y +\frac{27\sqrt{3}}{16\pi}\;
\Gamma^3(5/3)\;y^2 + O( y^3).
\eeq
At large $x$, $F_{N=0} (x=iy)$ grows exponentially with $y$, but still remains finite for all finite $y$.  We plot $F_{N=0} (iy)$ in  Fig.~\ref{fig:F0}.

\subsection{Nonanalytic terms in the self-energy}

Our key interest is how the pole disappears between $N=0$ and $N >>1$. One possibility might have been that the pole drifts to higher $x$ with decreasing $N$, and disappears at $x = i \infty$ at $N=0$. That would be consistent with the divergence of $F_{N=0} (iy)$ at $y = \infty$. However, the expansion in $1/N$ at large $N$ shows that the pole actually drifts to smaller $x =iy$ with decreasing $N$. Another possibility would be a behavior in which the residue $Z_N$
 of the pole gradually disappears at $N \rightarrow 0$, i.e., regular
corrections to $Z_N$ (the ones which are of order $\ln^2N/N^2$ at large $N$)
 make $Z_0 =0$. We, however, didn't find a self-consistent
 solution for $Z_N$ and $D_N$
 at small $N$ (this search required some computational  efforts).

Below we assume that regular corrections to fermionic propagator leave $Z_N$ and $D_N$ finite at $N \rightarrow 0$, and  the pole is destroyed by ``universal''  non-perturbative terms in the loop expansion of the self-energy near the mass shell. These terms appear at every order in the loop expansion, are
 non-analytic, and come from fermions with the lowest energies. We first identify these terms at $N=0$, and then show how they get modified at finite $N$.

\subsubsection{N=0}

The two-loop diagram is presented in Fig.~\ref{fig:se2} contains momentum and frequency integrals.
At $N=0$, the curvature term that couples the two disappears (see Eq.~\ref{1sigma2-1}). The momentum and frequency integration in the self-energy
 then factorize, and the momentum integrals are straightforwardly evaluated leaving only non-trivial frequency integrals. The two-loop self-energy  becomes
\begin{widetext}
\begin{eqnarray}
 \label{1sigma2-2}
\Sigma_2(k,\omega_m) &=& i\omega_m^{2/3}
\omega_0^{1/3}\left(\frac{\sqrt{3}}{2\pi}\right)^2\left(\frac{4\pi}{3\sqrt{3}}\right)^2
\int_0^1\frac{d
\Omega_1}{\Omega_1^{1/3}}\int_{1-\Omega_1}^1\frac{d
\Omega_2}{\Omega_2^{1/3}}
\nonumber\\
&\times& \frac{1}{\Delta +
\left[(1-\Omega_1)^{2/3}+(1-\Omega_2)^{2/3}+(\Omega_1+\Omega_2-1)^{2/3}-1
\right] },
\end{eqnarray}
\end{widetext}
where $\Omega_1$ and $\Omega_2$ are two internal frequencies normalized by $\omega_m$, and, we remind,  $\Delta=1+ i \epsilon_k/(\omega^{2/3}_m \omega^{1/3}_0)$.
Note that the coupling ${\bar g}$ is fully absorbed into $\omega_0$, i.e.,
 for a generic $\Delta = O(1)$, $\Sigma_2 (k, \omega_m)$ (as well as self-energies of higher-loop order) is of order $\omega^{2/3} \omega^{1/3}_0$, with a prefactor of order one.

Expanding Eq.~(\ref{1sigma2-2}) to first order in $\Delta$, we only obtain regular
 contributions to $Z_0$ and to $D_0$. These contributions are perturbative in
 the sense that typical internal $\Omega_{1,2}$ are of order one, much larger than $\Delta\ll 1$ over which one expands. However, this doesn't go beyond first order -- a formal expansion to order $\Delta^2$ yields a divergent prefactor. On more
 careful look, we found that the next term beyond $\Delta$
is actually $\Delta^{3/2}$. This term comes from the regions $\Omega_1 << 1$,
$1- \Omega_2 = \Omega^\prime_2 <<1$ and $\Omega_2 << 1$,
$1- \Omega_1 = \Omega^\prime_1 <<1$. The contributions from both regions are equal, and we only focus on the first one, for which the
 denominator in Eq.~(\ref{1sigma2-2}) becomes $\Delta + (\Omega^\prime_2)^{2/3} + (\Omega_1 - \Omega^\prime_2)^{2/3}$. Rescaling $\Omega_1 = \Delta^{3/2} x$, $\Omega^{\prime}_2= \Delta^{3/2} y$ and multiplying the self-energy by a factor of $2$ to account for the two regions,
we re-write Eq.~(\ref{1sigma2-2}) as
\begin{eqnarray}
\label{1sigma2-21}
\Sigma_2(k,\omega_m)&=& 8 i\omega_m^{2/3}\omega_0^{1/3}\;\frac{\Delta^{3/2}}{9}\\
&\times&\!\!\int_{0}^{1/\Delta^{3/2}}\frac{d x}{x^{1/3}}
\int_0^{x} \frac{dy}{1 + y^{2/3} + (x-y)^{2/3}}.\nonumber
\end{eqnarray}
The largest  contribution to this integral is confined to the
 upper limit of the integration over $x$. This is a
 regular, perturbative term (internal frequency $(\Omega_1)^{2/3} =
\Delta x^{2/3} >> \Delta$).
However, we also found a contribution which is confined to $x = O(1)$, i.e., to internal energies of order $\Delta^{3/2}$.  For the integral in (\ref{1sigma2-21}) this contribution can be easily singled out -- it is given by
\begin{eqnarray}
\label{1sigma2-21_a}
\Sigma_2(k,\omega_m)\! &=&\! \Sigma_{2,reg} (k, \omega_m)\qquad\qquad\quad + \\
&+&8 i\omega_m^{2/3} \omega_0^{1/3} \frac{\Delta^{3/2}}{9}
\int_{0}^{\infty}\frac{d x}{x^{1/3}} \int_0^{x} dy
 \nonumber\\
 &\times &\!\!\frac{1}{(1 + y^{2/3} + (x-y)^{2/3})( y^{2/3} + (x-y)^{2/3})^2}.
 \nonumber
\end{eqnarray}
Evaluating the integral numerically, we obtain
\begin{equation}
\label{S2zeroN}
\Sigma_2(k,\omega_m)=2.69\; i\omega_m^{2/3} \omega_0^{1/3}\Delta^{3/2}.
\end{equation}

At the first glance, this term is smaller than $O(\Delta)$ and is irrelevant to the issue of pole disappearance. Let's, however, continue our analysis and
verify whether there are singular renormalizations of $\Sigma_2 (k, \omega_m)$ from three-loop and higher-order diagrams. We found that such renormalizations do exist and come from vertex corrections. In particular, for
 the universal contribution to $\Sigma_2$ from  $\Omega_1 << 1$,
$1- \Omega_2 = \Omega^\prime_2 <<1$,
 singular  renormalization comes from  corrections to the two spin-boson vertices which involve bosonic propagator with
the frequency $\Omega_2$. Each vertex correction contains a block made out of    two extra Green's functions
 and one interaction line. We verified that the corrections to the two vertices
 involving the propagator with $\Omega_2$ are the same, and each
 yields, after integrating over internal $\epsilon_k$
\begin{eqnarray}
&& \frac{{\bar g}}{(2\pi)^2v_{\text{\tiny{F}}}} \int_{\omega \Omega_1}^\omega \frac{d \Omega}{\Sigma_1 (\Omega)} \int \frac{dq q}{q^3 + \gamma \Omega} = \nonumber \\
&& \frac{2}{3} \int_{\Omega_1}^1 \frac{d z}{z} = \frac{2}{3} \ln{\frac{1}{\Omega_1}}.
\label{sa_1}
\end{eqnarray}
We see that vertex correction is logarithmic and large, if $\Omega_1 \ll 1$.
Typical $\Omega_1$  for the universal term in $\Sigma_2$ are of order $\Delta^{3/2}$, hence the leading vertex correction is $\ln\bigl(1/\Delta\bigr)$.
Observe that the prefactor is just a number -- the coupling ${\bar g}$ is canceled out by $(\gamma \omega_0)^{1/3} \propto {\bar g}$.
Evaluating higher-order corrections in the leading logarithmical
 approximation (in which we keep only the highest power of the logarithm
 at any order),
 we find that the full vertex
 $\Gamma$ becomes (see Fig.~\ref{fig:se2}c)
\bea
\Gamma &=& 1  + \ln\frac{1}{\Delta} + \frac{1}{2} \left(\ln\frac{1}{\Delta}\right)^2 + \frac{1}{6} \left(\ln\frac{1}{\Delta}\right)^3 + ... \nonumber \\
&&= e^{\ln\frac{1}{\Delta}} = \frac{1}{\Delta}.
\label{su_3}
\eea
This $\Gamma$ is the solution of the differential RG equation
\be
\frac {d \Gamma}{d L} = \Gamma, \;\; L = \ln{\frac{1}{\Delta}}.
\ee
The implication is that the vertex correction can be treated within RG technique, i.e.,  that the system is renormalizable.

Eq.~(\ref{su_3})
 is not exact, however, as there exist extra  terms of order $\ln \bigl(1/\Delta\bigr)$
from second-order and higher-order vertex corrections. Such terms come with prefactors $O(1)$, and  modify the exponent in (\ref{su_3}) to
\be
\Gamma \propto  \frac{1}{\Delta^a},
\label{y_2}
\ee
where $a = O(1)$. Because the exponent $a$ has contributions from all orders,
 we cannot compute it explicitly  within the loop expansion,
 nor verify explicitly that logarithmic series still exponentiate beyond the leading logarithmic approximation (i.e., that RG procedure leading to (\ref{y_2})
 is still valid).
 At the same time, the situation here is not different from a variety of other problems where RG treatment has been adopted based on
the leading logarithmic series,~\cite{AIM,acs,artem} and
we assume that Eq.~(\ref{y_2}) is valid without further reasoning.

  Combining Eq.~(\ref{y_2}) with Eq.~(\ref{S2zeroN}), we find that the universal part of the self-energy, dressed by logarithmic vertex renormalizations, behaves near the mass shell as
\be
\Sigma_{univ} \sim  i \omega_m^{2/3} \omega_0^{1/3} \Delta^{3/2-2a}.
\label{su_4}
\end{equation}
To agree with the bosonization formula, we have to set $a=3/4$, then
$\Sigma_{univ}$ becomes $\Delta$-independent, and the full Green's function
 behaves near the former mass shell as $G^{-1} (k, \omega_m) = i \omega^{2/3}_m \omega^{1/3}_0 \left(A + B \Delta + ...\right)$ where $A$ and $B$ are constants.

We now re-evaluate universal terms at $N \neq 0$ and see how the universal self-energy gets modified at a finite $N$.\\

\subsubsection{Finite $N$}

 At a finite $N$, momentum and frequency integrals do not decouple, and the computations become more involved. Still, the
  two momentum integrals can be evaluated exactly, at any $N$, so the remaining task is to properly estimate the frequency integrals. As at $N=0$, we expand in $\Delta$ and verify whether or not this expansion contains the non-perturbative,
 universal term.   We find that
 such term is again present, but at a finite $N$ and $\Delta < N^2$,
scales as $\Delta^{5/2}$ rather than $\Delta^{3/2}$. This term comes from the same range of internal frequencies as at $N=0$ -- one of the internal frequencies is small, another is close to external $\omega$. Using the same notations as at $N=0$, we obtain for the contribution from such region, with logarithmic accuracy
\begin{eqnarray}
\label{univ2}
&&\Sigma_2(k,\omega_m) = i \omega_m^{2/3}
\omega_0^{1/3}\frac{\Delta^{5/2}}{4\pi^2N^2} \\
&\times&\!\int_0^{1/\Delta^{3/2}}\!\frac{dx}{x} \int_0^{x}dy
\left(1+y^{2/3}+(x-y)^{2/3}\right)\nonumber \\
&\times&\!\left[\ln\left(\frac{N}{\sqrt{\Delta}}\right)+\ln\left(\frac{x^{1/3}}{1+y^{2/3}+(x-y)^{2/3}}\right)\right]^2.\nonumber
\end{eqnarray}
The  universal term (the one which does not depend on the upper limit of integration) comes from the cross-product of the two logarithms, and the universal part of the two-loop self-energy is
\begin{eqnarray}
\label{sigma2N}
\Sigma_{2} (k, \omega_m) = i \omega_m^{2/3}
\omega_0^{1/3} \Delta^{5/2} \frac{J}{8 \pi^2 N^2}~ \ln{\frac{N^2}{\Delta}}.\;\;\;\;\;\;\;
\end{eqnarray}
The prefactor $J$ is given by
\be
 J = - \int_0^\infty dx \int_0^1 dz \psi (x,z),
\ee
where
\begin{eqnarray}
\label{su_8}
 \psi (x,z)\!\! &=&\!\! \ln \left(1 + \frac{1}{x^{2/3} f_z}\right) -
\frac{1}{x^{2/3} f_z}
+x^{2/3} f_z\\
&\times&\!\!\left[\ln\left(1 + \frac{1}{x^{2/3} f_z}\right) -
\frac{1}{x^{2/3} f_z} + \frac{1}{2}\left(\frac{1}{x^{2/3} f_z}\right)^2\right],\nonumber
\end{eqnarray}
and $f_z = z^{1/3} + (1-z)^{1/3}$.
Evaluating the integral, we obtain $J \approx 0.96$.

The next step is to include vertex renormalizations from higher-order diagrams, i.e., renormalize the universal part of $\Sigma_2$ in (\ref{sigma2N}) into
 $\Sigma_{2,univ} =\Sigma_{2} * \Gamma^2$. We found that multi-loop contributions to $\Gamma$ are again  logarithmic, but now the lower limit of the logarithm is set by the largest of $\Delta$ and $N^2$. The full result is then
$\Gamma \propto 1/\Delta^{3/4}$ for $\Delta \gg N^2$ and $\Gamma \propto 1/N^{3/2}$ for $\Delta << N^2$. Combining this with (\ref{sigma2N}), we obtain
\begin{eqnarray}
\label{sigma2N_1}
\Sigma_{univ} \sim  i \omega_m^{2/3}
\omega_0^{1/3} g \left(\frac{\Delta}{N^2}\right),
\end{eqnarray}
where $g(y <<1) \propto y^{5/2}$ and $g(y >>1) = {\text const}$.
We neglected $\ln{\Delta/N^2}$ term because $\Gamma$ is only known to the accuracy that neglects possible additional logarithmical factors.

Substituting  the full $\Sigma_{univ}$ into the Green's function,
we reproduce Eq.~(\ref{y_5}):
\bea
&&G(k, \omega_m) = -\frac{1}{\epsilon_k} F_{N \rightarrow 0} \left(\Delta\right); \nonumber \\
&& F_{N \rightarrow 0} \left(\Delta\right) = \frac{1}{-\Delta + g \left
(\Delta/N^2\right)}.
\label{y_51}
\eea
 We see that the pole survives, at the smallest $\Delta$ for any nonzero $N$ and, moreover, its residue remains $Z_N = O(1)$. However,  at distances from the pole larger than $N^2$, the system evolves, and its
behavior becomes essentially the same as at $N=0$.

\section{Half-filled Landau level}

In this section we study  different example of a non-Fermi liquid behavior -
 2D fermions at half-filled Landau level with unscreen Coulomb interaction.~\cite{Jain,Halperin,AIM} The effective low-energy theory is described by
 Eq.~(\ref{Act1}) with massless bosonic propagator
\begin{eqnarray}
\label{khi1} \chi(q,\omega)=\frac{\chi_0}{mu
q+\gamma\;|\omega/q|}.
\end{eqnarray}
Here $u$ is the effective velocity [$u=e^2/(8\pi{\hat\epsilon})$ with ${\hat\epsilon}$ the
dielectric constant of a host semiconductor], and $\gamma
= {\bar g} k_{\text{\tiny{F}}}/(\pi v^2_{\text{\tiny{F}}})$
 is the same as in the $\omega^{2/3}$
problem.

We start by presenting the results
for the limiting cases $N \rightarrow \infty$ and $N=0$.
Large $N$ limit is studied in $1/N$
expansion while  $N=0$ is
obtained via 1D bosonization.  We then discuss our results for the
universal terms in the loop expansion near the mass-shell.


\subsection{Fermionic propagator at $N>>1$}

The one-loop self-energy
 has been calculated in Ref.~\onlinecite{AIM}.
It  has the form
\begin{eqnarray}
\label{log1}
\Sigma_1(\omega_m)=i\lambda\;\omega_m\ln
\biggl(\frac{\epsilon_0}{|\omega_m |}\biggr),
\end{eqnarray}
where $\lambda=\bar{g}/\left(4\pi^2uk_{\text{\tiny{F}}}\right)$
 is a dimensionless coupling and $\epsilon_0$ is an energy scale defined by $\epsilon_0=(2\pi/N)\left(E_{\text{\tiny{F}}}/\bar{g}\right)uk_{\text{\tiny{F}}}$.
The momentum dependent part of $\Sigma$ is regular and  we neglect it.
 The $\omega \log \omega$ form of the self-energy Eq.~(\ref{log1}) implies that
the system exhibits a marginal non-Fermi liquid behavior, however the Fermi surface remains well defined.

As in the previous case, higher order self-energy diagrams are
parametrically small in $(\ln N/N)^2$.
Near the mass-shell $\tilde{\Delta}=0$, where
\begin{eqnarray}
\label{Dtilde}
\tilde{\Delta}=1+\frac{i\epsilon_k}{\lambda\omega_m\ln\left(\epsilon_0/\omega_m\right)},
\end{eqnarray}
 we found at two-loop order (see Appendix~\ref{CS}) that
$\Sigma_2 (k, \omega_m)$ is linear in $\tilde{\Delta}$
\begin{eqnarray}
\label{MargSigma2}
\Sigma_2 (k,\omega_m)=i\lambda\omega_m\ln \left(\epsilon_0/\omega_m\right)
 \frac{\tilde \Delta}{24} \left(\frac{\ln N}{N}\right)^2.
\end{eqnarray}
We see that, as for the $\omega^{2/3}$ problem,  multi-loop self-energy terms acquire
extra powers of $(\ln N/N)^2$.
Substituting one- and two-loop self energies into the
Green's function and rewriting it as in Eqs.~(\ref{n_1}), (\ref{n_3}), we obtain $G({\bf k}, \omega_m)=-\left(1/\epsilon_k\right)\tilde{F}_N(-y)$, where $y=\left(\Sigma_1/i\epsilon_k\right)$, and
 $\tilde{F}_{N\rightarrow\infty}(y)$ is again given by Eq.~(\ref{n_3}),
 but now
\begin{eqnarray}
\label{Zfactor}
Z_N = 1-\frac{1}{24}\;\left(\frac{\ln N}{N}\right)^2, ~~D_N =1.
\end{eqnarray}
 We see that  $1/N$ corrections for the Landau level case
 are qualitatively similar to the case of a nematic QCP -- in both cases, the pole in $G({\bf k}, \omega_m)$ viewed as a function of $\epsilon_k$
survives, but acquires a residue $Z_N <1$.

In real space, we have (see Appendix~\ref{z=2})
\begin{eqnarray}
\label{compare}
&&G({\bf r})=G_0({\bf r})\left[\frac{1}{\ln(r/r_0)}\right], v_F t < r \ln {r/r_0} ,\nonumber \\
&&G({\bf r},t) \approx G_0({\bf r},t),\;\;\;\;v_{\text{\tiny{F}}}t >
r\ln (r/r_0),
\end{eqnarray}
where $G_0({\bf r},t)\propto (r+v_{\text{\tiny{F}}}t)^{-1}\sqrt{k_{\text{\tiny{F}}}/r}$ is a propagator of a free fermion, and
 $r_0=v_{\text{\tiny{F}}}/\epsilon_0$.


\subsection{Low-energy effective theory for $N=0$: Fermionic propagator}

As in the previous case, at $N=0$ the curvature of the Fermi surface
becomes unimportant,   the motion of fermions becomes essentially
one-dimensional, and the fermionic propagator can be obtained
 by mapping the original 2D problem  to an effective 1D theory.
 \cite{Lidsky,AIM} For a half-filled Landau level
the effective action has the form
\begin{eqnarray}
\label{F-action} {\mathcal S}=\int\!d\Omega
dk\qquad\qquad\qquad\qquad\qquad\qquad\qquad\qquad\;\\
\times\Bigl[\bar\psi_{-\Omega,-k}^{R,a}\bigl(i\Omega-v_{\text{\tiny{F}}}k\bigr)\psi_{\Omega,k}^{R,a}
+\bar\psi_{-\Omega,-k}^{L,a}\bigl(i\Omega+v_{\text{\tiny{F}}}k\bigr)\psi_{\Omega,k}^{L,a}\nonumber\\
+\left(4\pi^2\lambda v_{\text{\tiny{F}}}\right)\ln\Omega\;\vert\rho^{R}_{\Omega,k}-\rho^{L
}_{\Omega,k}\vert^2\Bigr],\nonumber
\end{eqnarray}
where $\rho_{\omega,k}=\bar\psi_{\omega,k}^a\psi_{\omega,k}^a$ is
the density operator, and the replica index $a$ has $s$ values. As usual,
 we take the limit $s\rightarrow 0$
in order to avoid
generating fermionic loops in perturbation theory.

We  solve the
model Eq.~(\ref{F-action}) by 1D bosonization. Following the
standard steps we find that
\begin{eqnarray}
\label{1DG}
G^{1D}(r,\tau)=\frac{i}{2\pi(r-iv_{\text{\tiny{F}}}\tau)}
\exp\left[-\lambda\;\frac{r\ln |r-iv_{\text{\tiny{F}}}\tau|}{(r-iv_{\text{\tiny{F}}}\tau)}\right],
\nonumber\\
\end{eqnarray}
where  $\tau$ is Matsubara time.
A Fourier transform of Eq.~(\ref{1DG})
yields a simple scaling form for $G({\bf k},\omega_m)$:
\begin{eqnarray}
\label{Gscale}
G({\bf k},\omega_m)\!&=&\!-\frac{1}{\epsilon_k}\tilde{F}_{N=0}
\left[\frac{-\lambda\omega_m\ln
\left({\epsilon_0}/{|\omega_m |}\right)}{\epsilon_k} \right],\nonumber\\
\tilde{F}_{N=0}(x)\!&=&\!e^{-ix}.
\end{eqnarray}
We verified that the loop expansion  reproduces the
small $x$ expansion of $\tilde{F}_{N=0}(x)$. Obviously,
 $\tilde{F}_{N=0} (x)$ has no poles along $x = i y$ where $\tilde{F}_{N=0} (iy ) = e^y$. We verified that the  singularity at $ y\rightarrow\infty$
 is responsible for the exponential  behavior of $G({\bf r},t)$ in
 (\ref{1DG}).

\vspace{1cm}
\subsection{Nonanalytic terms in the self-energy}

\subsubsection{ $N=0$}
The relevant two-loop self-energy is again given by the diagram
 in Fig.~\ref{fig:se2}a. We compute it in  Appendix~\ref{CS}
for arbitrary $N$.  As in the previous case, consider first $N=0$.
We have
\begin{widetext}
\begin{eqnarray}
\label{CS0N}
&&\Sigma_2(k,\omega_m)= i\lambda\;\omega_m\int_{0}^{1}
d\Omega_1\int_{1-\Omega_1}^{1}d\Omega_2
\\
&&\frac{\ln\left[\epsilon_0/(\omega_m\Omega_1)\right]\;\ln\left[\epsilon_0/(\omega_m\Omega_2)\right]}{
\tilde{\Delta}\ln\left(\epsilon_0/\omega_m\right)+
\left\{(1-\Omega_1)\ln\left[\frac{1}{1-\Omega_1}\right]+(1-\Omega_2)\ln\left[\frac{1}{1-\Omega_2}\right]
+(\Omega_1+\Omega_2-1)\ln\left[\frac{1}{\Omega_1+\Omega_2-1}\right]\right\}
},\nonumber
\end{eqnarray}
\end{widetext}
 where $\Omega_1$ and $\Omega_2$ are the two internal energies normalized by $\omega_m$. In the previous case, $\Sigma_2$ was regular to first order in $\Delta$.  This time, expanding Eq.~(\ref{CS0N})  to first order in
$\tilde{\Delta}$ we found that the prefactor diverges logarithmically.
 This suggests that the term next to the constant
 is already nonanalytic. We explicitly verified that the
nonanalyticity emerges in Eq.~(\ref{CS0N})
 from the regions $\Omega_1 \ll 1$,
$\Omega_2\sim 1$ and $\Omega_2 \ll 1, \Omega_1 \sim 1$.
 Within the logarithmic accuracy, the nonanalytic contribution to $\Sigma_2$  can then be cast into the form
\begin{eqnarray}
\label{CS3}
&&\Sigma_2(k,\omega_m)= -2 i\lambda \omega_m \; \tilde{\Delta}\;\ln^2\left(\epsilon_0/\omega_m\right)\nonumber\\
&&\times \int_{\frac{{\tilde \Delta} \ln{\left(\epsilon_0/\omega_m\right)}}{|\ln{\tilde\Delta}|}}^1 \frac{d\Omega_1}{\Omega_1 \ln {1/\Omega_1}} \nonumber \\
&& =-2 i\lambda\;\omega_m {\tilde \Delta}\;\ln^2\left(\epsilon_0/\omega_m\right)\;\ln\ln\left(\frac{1}{\tilde{\Delta}}\right).
\end{eqnarray}

As before, higher-order diagrams modify Eq.~(\ref{CS3}) by adding vertex corrections. A building block for a vertex correction is again the product of  two
fermionic Green's functions and one interaction line. Evaluating this block
we find  that vertex corrections are again logarithmical.
 Using further
  the fact that typical dimensionless external energy for the
 correction to the vertex involving a boson with $\Omega_2$
 is of order $\Omega_1$,
we obtain the renormalized vertex ${\tilde \Gamma}$ in the form
\be
{\tilde \Gamma}  = 1 + \ln\left(1/\Omega_1\right).
\ee
Evaluating further higher-order vertex corrections, we obtain in the leading logarithmical approximation
\begin{eqnarray}
\label{vertCS}
{\tilde \Gamma} &=&1+\ln{\frac{1}{\Omega_1}} +
\frac{1}{2} \left(\ln{\frac{1}{\Omega_1}}\right)^2+\ldots \nonumber \\
&& = e^{\ln{1/\Omega_1}} = \frac{1}{\Omega_1}.
\end{eqnarray}
As in the previous case, the exponent is modified by higher-order corrections to
\be
{\tilde \Gamma} \sim \left(\frac{1}{\Omega_1}\right)^b.
\label{y_7}
\ee
A simple experimentation shows that the full universal self-energy $\Sigma_{univ} = \Sigma_2 {\tilde \Gamma}^2$ agrees with bosonization if we set $b=1/2$. Indeed, in this case, we have, combining Eqs.~(\ref{y_7}) and (\ref{CS3})
\begin{eqnarray}
\label{S2univ}
&&\Sigma_{univ}= -i c \lambda\;\omega_m {\tilde \Delta} \ln^2\left(\epsilon_0/\omega_m\right) \times \\
&&\int_{\frac{{\tilde \Delta} \ln{\left(\epsilon_0/\omega_m\right)}}{|\ln{\tilde\Delta}|}}^1
\frac{d\Omega_1}{\Omega^{1+2b}_1 \;[\ln {1/\Omega_1}]}  = -i c \lambda \omega_m \ln \left(\epsilon_0/\omega_m\right),\nonumber
\end{eqnarray}
where $c = O(1)$. We see that the fully renormalized $\Sigma_{univ}$  tends to a finite value on the mass shell, i.e.,
\begin{equation}
G^{-1} \bigl(\tilde{\Delta}\bigr) = i \lambda \omega_m \ln \left(\epsilon_0/\omega_m\right) \left[{\tilde \Delta} + c\right].
\label{d_1}
\end{equation}
This behavior is consistent with the bosonization formula, Eq.
(\ref{Gscale})  which yields  $G^{-1} \propto (1 + {\tilde \Delta})$
 near the mass shell. The computation of the constant $c$ is beyond the scope of our analysis.

\subsubsection{Finite $N$}

We now check how the expression for the self-energy is
 modified at finite $N$.  Just like at $N=0$, we
expand  the two-loop self-energy in powers of $\tilde{\Delta}$
and extract the  universal term in the prefactor.
We find that, to logarithmical accuracy, the linear in ${\tilde \Delta}$ term
 is already universal, like at $N=0$, and
 the only difference between $N=0$ and $N >0$ cases is in lower
 limit of the frequency integration over $\Omega_1$: at a finite $N$, instead of ${\tilde \Delta}$, as in (\ref{CS3}), we now have $N^2$.
Accordingly, instead of ((\ref{CS3}), we now have
\begin{eqnarray}
\label{CSS2}
\Sigma_2(k,\omega_m)&=&-i\lambda\;\omega_m\;
\ln^2\left(\epsilon_0/\omega_m\right)\;\tilde{\Delta}\ln\ln(1/N).\nonumber\\
\end{eqnarray}

As the next step, we include vertex corrections.
They are still logarithmical and exponentiate, as in (\ref{y_7}), but now
typical $\Omega_1$ are of order $N^2$, hence
\begin{eqnarray}
\label{FNS2univ}
\Sigma_{univ} \propto -i\lambda\;\omega_m\ln^2\left(\epsilon_0/\omega_m\right)
\;\frac{\tilde{\Delta}}{N^2}.
\end{eqnarray}
We neglected $\ln |\ln N|$ factor as the renormalized vertex is only known to this accuracy. We  see that $\Sigma_{univ}$ is still linear in $\Delta$, i.e., the pole in the Green's function survives in the range of order $N^2$.
This is similar to $\omega^{2/3}$ case. Unlike that case, however,  the
 residue of the pole scales as $N^2$ and vanishes when $N =0$.
Substituting  $\Sigma_{univ}$ into the Green's function and
rewriting the result in terms of $\tilde F_N$, as in (\ref{Gscale}), we
 obtain near the mass shell
\begin{eqnarray}
\label{Gscale_1}
&&G({\bf k},\omega_m)\!= \!-\frac{1}{\epsilon_k}\tilde{F}_{N \rightarrow 0}
\left({\tilde \Delta}\right),\nonumber\\
&&\tilde{F}_{N \rightarrow 0} \left({\tilde \Delta}\right) = {\tilde F} \left(\frac{\tilde\Delta}{N^2}\right),
\end{eqnarray}
where ${\tilde F}(y \ll 1) \propto y, \: {\tilde F}(y \gg 1) = {\text const}$.

The difference in $Z_N$ between the QCP and the Landau level cases is due to the fact that in  the case of a Landau level  the
 universal self-energy already emerges in the loop expansion
already at  order ${\tilde \Delta}$ while at a nematic QCP
it appears at order $\Delta^{3/2}$.
 There may also be additional differences between the scaling functions
$F_{N\rightarrow 0}$ and ${\tilde F}_{N \rightarrow 0}$ due to extra logarithmic or doubly logarithmic factors, but, as we said, these factors are
  beyond the slope of our paper.

\section{conclusions}
In this paper we considered two examples of
 2D fermions coupled to critical overdamped bosonic fields.
One  describes a nematic and an Ising ferromagnetic quantum critical points,
 other describes a half-filled Landau level. For both cases, the low-energy physics is governed by the interaction between fermions and  collective neutral excitations
 with the static propagator $\chi (q) = \chi_0/q^{1+x}$.  For a nematic and ferromagnetic QCP  $x=1$,
 for a half-filled Landau level with unscreened Coulomb interaction, $x=0$.
 In both cases, quantum fluctuations destroy a coherent Fermi liquid behavior down to the lowest energies, but leave the Fermi surface intact. The issue we addressed is what is the form of the fermionic propagaror, and whether it has a pole as a function of the dispersion $\epsilon_k$.

The low energy properties of these systems can be studied in a controllable way
by extending the theory to $N >>1$ fermionic
flavors and assuming that the interaction with neutral excitations conserves the flavor. At large $N$, self-energy is perturbative in $1/N$, and the
 pole in the Green's function survives despite that $\Sigma (k_{\text{\tiny{F}}}, \omega)
 \propto \omega^{2/3}$ for $x=1$ and $\Sigma (k_{\text{\tiny{F}}}, \omega) \propto \omega \log \omega$ for $x=0$. $1/N$ corrections only affect the residue of the pole $Z_N$ and make $Z_N <1$.  Existence
of the pole implies that the propagator in real space/time is long ranged and decays by a power-law, e.g., as
$G(r\gg v_{\text{\tiny{F}}}t)\propto1/r^2$ and $G(r\ll v_{\text{\tiny{F}}}t)\propto1/t\sqrt{r}$ for $x=1$. The other limit  $N=0$, has been solved using
 1D bosonization, and the result is that for both $x=1$ and $x=0$, fermionic propagator does not have a pole, and the propagator $G(r,t)$ is short-ranged.

The issue we  addressed is at what $N$ the pole disappears.
This is essential as the physical case $N=1$ is ``in between'' the two limits.
We performed the loop expansion for the self-energy both at $N=0$ and at a finite but small $N$. At $N=0$, we identified universal, non-analytic
 contributions to $\Sigma$ which destroy the pole and make fermionic propagator regular near the former mass shell. At small but nonzero $N$, we found that singular, universal
 terms in the self-energy still exist, but they do not destroy the pole
 in the range of order $N^2$ around the mass shell.
 At larger deviations from the mass shell, fermionic propagator recovers the same regular form as at $N=0$.
 For a nematic QCP, the residue of the pole $Z_N$ remains $O(1)$ in this range, while for the case of a half-filled Landau level, the residue of the pole scales as $Z_N \propto N^2$.

The result that the pole in $G(k, \omega_m)$ exists for any $N>0$ and only vanishes at $N=0$ agrees with the conjecture by Altshuler {\em et al}~\cite{AIM}.
The key result of our work is the identification of peculiar, universal terms in the loop expansion of the self-energy,  which are responsible for the destruction of the pole at $N=0$. At $N >0$, these terms get modified -- they do not destroy the pole but reduce the  width where the pole exists.

\section{acknowledgement}
We acknowledge with thanks useful conversations with E.~Fradkin, D.~Khveshchenko, Y.~B.~Kim, M.~Lawler, W.~Metzner, and T.~Senthil.  The work was supported by
  NSF-DMR 0604406 (A.~V.~C) and by Herb foundation (T.~A.~S).


\appendix

\section{Analysis of the fermion propagator at a QCP
at $N\rightarrow\infty$ and $N=0$}
\label{pole}
In this Appendix we analyze the behavior of the fermion propagator
\begin{eqnarray}
\label{Grt}
G({\bf r},\tau)=\int_{-\infty}^{\infty}\frac{d\omega_m}{2\pi}\int\frac{d{\bf k}}{(2\pi)^2}\;e^{i{\bf k}{\bf r}-i\omega_m \tau}G({\bf k},\omega_m),
\end{eqnarray}
 at a nematic QCP at $N>>1$ and $N=0$. The goal of this analysis is to demonstrate that the form of long-distance behavior of $G({\bf r},\tau)$, where $\tau$ is the Matsubara time, is qualitatively different depending on whether or not $G({\bf k}, \omega_m)$ has a pole as a function of $\epsilon_k$.

We use the Matsubara frequency form of $G(\epsilon_k, \omega_m) = -(1/\epsilon_k) F_N \left(-\omega_m^{2/3} \omega^{1/3}_0/\epsilon_k\right)$.
Introducing new variables,
 $x=-(\omega_m^{2/3} \omega^{1/3}_0)/\epsilon_k$,
$y=(\epsilon_kr/v_{\text{\tiny{F}}})$ and $\alpha=\left(v_{\text{\tiny{F}}}^{3/2}\tau/\omega_0^{1/2}r^{3/2}\right)$
we re-write  Eq.~(\ref{Grt}) as
 \begin{eqnarray}
 \label{INform}
 G({\bf r},\tau)\!\! &=&\!\! \frac{3}{2r^2} \left(\frac{1}{\pi^{3/2}}\right)\sqrt{\frac{E_{\text{\tiny{F}}}}{\omega_0}}
 \int_{-\infty}^\infty dx|x|^{1/2}\!\int_{-\infty}^{\infty}\!dy|y|^{1/2}\nonumber\\
 &\!\times &\!\cos\left(k_{\text{\tiny{F}}}r+y-\frac{\pi}{4}\right)
F_{N}(x)e^{-i(-1)^{3/2}x^{3/2}y^{3/2}\alpha}.\nonumber\\
 \end{eqnarray}
\subsection{Limit $N\rightarrow\infty$}

In this limit $F_{N=\infty}(x)=1/(1+ix)$.
Substituting this into Eq.~(\ref{INform}), we obtain
 \begin{eqnarray}
 \label{Iform}
 G({\bf r},\tau)&=&\frac{1}{r^2} \left(\frac{3}{2\pi^{3/2}}\right)\sqrt{\frac{E_{\text{\tiny{F}}}}{\omega_0}}
 \int_{-\infty}^{\infty}dx\int_{-\infty}^{\infty}dy\;|xy|^{1/2}\nonumber\\
 &\times&\frac{e^{-i(-1)^{3/2}x^{3/2}y^{3/2}\alpha}}{1+ix}\cos\left(k_{\text{\tiny{F}}}r+y-\frac{\pi}{4}\right).
 \end{eqnarray}
 We now show that the pole at $x=-i$ determines long-distance behavior of
$ G({\bf r},t)$. For definiteness, we focus on the case  $\alpha <<1$, i.e.,
$v_{\text{\tiny{F}}}^{3/2}\tau << \omega_0^{1/2}r^{3/2}$.
 In this limit,  the exponential factor in
(\ref{Iform}) can be set to unity and the integrals
 decouple. Integration over $y$ is straightforward and yields
\begin{eqnarray}
\label{reguly}
\lim_{\lambda\rightarrow 0^+}\int_{-\infty}^{\infty}dy|y|^{1/2}\cos\left(k_{\text{\tiny{F}}}r+y-\frac{\pi}{4}\right)e^{-\lambda |y|}\nonumber\\
=-\sqrt{\frac{\pi}{2}}\cos\left(k_{\text{\tiny{F}}}r-\frac{\pi}{4}\right).
\end{eqnarray}
We added the factor $e^{-\lambda |y|}$ for regularization.
The subsequent integration over $x$ is also straightforward.
\begin{eqnarray}
\label{intx}
\int_{-\infty}^{\infty}dx\frac{|x|^{1/2}}{1+ix}=2\int_{0}^{\infty}dx\frac{x^{1/2}}{1+x^2}=\sqrt{2}\pi.
\end{eqnarray}
We also evaluated the
integral in Eq.~(\ref{intx}) by extending the integration into
 the complex plane of $x$ and found that the result comes from the pole at
$x=i$, the branch cut contribution in (\ref{intx}) is cancelled out.
 Substituting Eqs.~(\ref{reguly}) and (\ref{intx}) into Eq.~(\ref{Iform})
 we find that the fermion propagator
$G({\bf r},\tau)\propto 1/r^2$ at $v_F\tau\ll r^{3/2}/r_0^{1/2}$, where $r_0=v_F/\omega_0$. We emphasize again that the power-law decay is the consequence of the pole in $G({\bf k}, \omega_m)$.

For completeness, we also present the result for  $G({\bf r},\tau)$
for $\alpha\gg 1$ . In this case we obtained
 \begin{eqnarray}
\label{withg}
 G({\bf r},\tau)=\frac{1}{r^2} \left(\frac{3}{\sqrt{2\pi}}\right)\sqrt{\frac{E_{\text{\tiny{F}}}}{\omega_0}}\;{\cal K}(\alpha),
\end{eqnarray}
where
 \bea
 \label{KK}
 {\cal K}(\alpha)\text{\Large{$|$}}_{\alpha\rightarrow\infty}=
 \frac{2\sqrt{2}}{3\alpha}\cos\left(k_{\text{\tiny{F}}}r-\frac{\pi}{4}\right).
 \ea
Using $\alpha \propto \tau/r^{3/2}$ we find that in this limit the
real space/time Green's function behaves as $G({\bf r},\tau)\propto 1/\left(\sqrt{r}v_{\text{\tiny{F}}} \tau\right)$.
Combining the two results, we have
\begin{eqnarray}
\label{NMN}
G({\bf r},\tau)\text{\Large{$|$}}_{N\rightarrow\infty}\propto\left\{
  \begin{array}{c}
    \frac{1}{r^2},\;\;\; v_{\text{\tiny{F}}} \tau \ll r^{3/2}/r_0^{1/2}\\
    \frac{k_{\text{\tiny{F}}}}{v_{\text{\tiny{F}}} \tau}\sqrt{\frac{r_0}{r}},\;\;\; v_{\text{\tiny{F}}} \tau \gg r^{3/2}/r_0^{1/2}\\
  \end{array}
\right. ,
\end{eqnarray}
or, in terms of the free-fermion propagator
$G_0({\bf r},\tau)$,
\begin{eqnarray}
\label{NMN_1}
G({\bf r},\tau)\text{\Large{$|$}}_{N\rightarrow\infty}=\left\{
  \begin{array}{c}
    G_0({\bf r},\tau)\sqrt{r_0/r},\;\;\; v_{\text{\tiny{F}}} \tau\ll r^{3/2}/r_0^{1/2}\nonumber\\
    G_0({\bf r},\tau)\sqrt{k_{\text{\tiny{F}}}r_0},\;\;\; v_{\text{\tiny{F}}} \tau \gg r^{3/2}/r_0^{1/2}\nonumber\\
  \end{array}
\right. .\\
\end{eqnarray}

\subsection{Limit $N=0$}
The expression for $G(k, \omega_m)$ in this case is given
 in Ref.~\onlinecite{AIM}, see Eq.~(\ref{scaling2}).
The function $F_{N=0}(x)$ in (\ref{scaling2}) can be represented as
\begin{eqnarray}
\label{standard}
F_{N=0}(x)&=&\cosh\left[e^{i\pi/4} \Gamma^{3/2}(5/3) x^{3/2}\right]-ix\;{\mathcal F}_1(x)\nonumber\\
&-&x^2\frac{27\sqrt{3}}{16\pi}\Gamma^3(5/3) \;{\mathcal F}_2(x),
\end{eqnarray}
where ${\mathcal F}_1$ and ${\mathcal F}_2$ are expressed
 in terms of hypergeometric functions
\begin{eqnarray}
\label{hyper}
{\mathcal F}_1(x)=\text{\tiny{1}}\large{F}\text{\tiny{2}}\left[1;5/6,4/3;(1/4)i\Gamma^{3}(5/3)\;x^3\right],\nonumber\\
{\mathcal F}_2(x)=\text{\tiny{1}}\large{F}\text{\tiny{2}}\left[1;7/6,5/3;(1/4)i\Gamma^{3}(5/3)\;x^3\right].
\end{eqnarray}
At small arguments, ${\mathcal F}_1(x)$ and ${\mathcal F}_2(x)$ are expandable in $x$~\cite{AIM}
\begin{eqnarray}
\label{expansions}
{\mathcal F}_1(x)\!=\!1+\frac{9}{40}i\left[\Gamma(5/3)x\right]^3-\frac{81}{6160}\left[\Gamma(5/3)x\right]^6+O\left[ix^9\right],\nonumber\\
{\mathcal F}_2(x)\!=\!1+\frac{9}{70}i\left[\Gamma(5/3)x\right]^3-\frac{81}{14560}\left[\Gamma(5/3)x\right]^6+O\left[ix^9\right].\nonumber\\
\end{eqnarray}
By contrast,  at large $|x|\gg 1$, $F_{N=0} (x)$ diverges as
\begin{eqnarray}
\label{largex}
F_{N=0}(x)&=&\frac{3}{2}\exp\left[-e^{i\pi/4}\Gamma^{3/2}(5/3)x^{3/2}\right]\\
&+&\left[-\frac{3i\sqrt{3}}{4\pi x}-\frac{2}{9\Gamma^{3}(5/3)x^2}-O\left[1/x^4\right]\right].\nonumber
\end{eqnarray}
Still, along  $x=iy$
$F_{N=0}(iy)$ as a regular function of $y$ for all finite $y$, i.e., there is no  pole.
The plot of  $F_{N=0}(iy)$ versus $y$ is presented in Fig. \ref{fig:F0}.

Below we show that the absence of the pole eliminates a
power-law decay of $G(r,\tau)$, while the divergence of $F_{N=0} (x)$ at
infinity gives rise to  the
exponential decay  of the fermionic propagator.

The easiest way to evaluate the integral over $x$
in Eq.~(\ref{INform})  with $F_{N=0}(x)$
given by Eq.~(\ref{standard}) is to deform the integration contour into the complex plane. As for large $N$, the contribution from the branch cut is
canceled out and in the absence of the pole $G(r,\tau)$ only comes from the integral over a semi-circle with infinite radius. Using parametrization
\begin{eqnarray}
\label{parametrization}
x^{3/2}=R^{3/2}\exp\left\{i\varphi\right\},\;\;\;\;\pi<\varphi<2\pi,
\end{eqnarray}
with $R\rightarrow \infty$, substituting into (\ref{INform}) the
large-$x$ asymptote of $F_{N=0}$
 \begin{eqnarray}
\label{asymptote}
F_{N=0}(x)=\frac{3}{2}\exp\left[-e^{i\pi/4+i\varphi}\Gamma^{3/2}(5/3)R^{3/2}\right],
\end{eqnarray}
and introducing $y_1=\alpha^{2/3}y$, we obtain
\begin{eqnarray}
\label{overphi}
G({\bf r},\tau)=-\frac{1}{\alpha
r^{2}} \left(\frac{3}{2\pi^{3/2}}\right)\sqrt{\frac{E_{\text{\tiny{F}}}}{\omega_0}}
 \int_{-\infty}^{\infty}dy_1|y_1|^{1/2}\nonumber\\
\times\cos\left(k_{\text{\tiny{F}}}r+\left[y_1/\alpha^{2/3}\right]-\frac{\pi}{4}\right)
\int_{\pi}^{2\pi}\!d\varphi \;iR^{3/2}e^{i\varphi}\nonumber\\
\times\frac{3}{2}\exp\left\{R^{3/2}e^{i\varphi}\left[-iy_1^{3/2}-e^{i\pi/4}\Gamma^{3/2}(5/3)\right]\right\}.\nonumber\\
\end{eqnarray}
Integrating over the angular variable $\varphi$ we obtain
\begin{eqnarray}
\label{int1} R^{3/2}\int_{\pi}^{2\pi}d\left\{
e^{i\varphi}\right\}\exp\Biggl\{R^{3/2}
e^{i\varphi}\qquad\;\;
\nonumber\\
\qquad\times\left[-iy_1^{3/2}-e^{i\pi/4}\Gamma^{3/2}(5/3)\right]\Biggr\}\nonumber\\
={2\pi}\;\delta\left[iy_1^{3/2}+e^{i\pi/4}\Gamma^{3/2}(5/3)\right],\quad
\end{eqnarray}
where $\delta (...)$ is the $\delta$-function.
The integration over $y_1$ is then straightforward and yields
\begin{eqnarray}
\label{GG1}
&&G({\bf r},\tau)=-\frac{1}{\alpha
r^{2}} \left(\frac{3}{\pi^{1/2}}\right)\sqrt{\frac{E_{\text{\tiny{F}}}}{\omega_0}}
 \int_{-\infty}^{\infty}dy_1|y_1|^{1/2}\times\nonumber\\
&&\cos\left(k_{\text{\tiny{F}}}r+\left[y_1/\alpha^{2/3}\right]-\frac{\pi}{4}\right)
\delta\left[iy_1^{3/2}+e^{i\pi/4}\Gamma^{3/2}(5/3)\right]\nonumber\\
&&\propto\frac{1}{\tau\sqrt{r}}\exp\left\{-\frac{\Gamma(5/3)}{(i)^{2/3}\alpha^{2/3}}\right\}.
\end{eqnarray}
Re-expressing $\alpha$ in terms of $r$ and $t$, we finally obtain
\begin{eqnarray}
\label{logProp-3} G({\bf r},\tau)\propto \frac{k_{\text{\tiny{F}}}}{v_{\text{\tiny{F}}} \tau}\sqrt{\frac{r_0}{r}}
\exp\Biggl(-\frac{\Gamma(5/3)r}{r_0^{1/3} \bigl[i v_{\text{\tiny{F}}}
\tau\bigr]^{2/3}}\Biggr).
\end{eqnarray}
We see that $G(r,\tau)$ is now exponential. The full bosonization result contains $(iv_{\text{\tiny{F}}}\tau-r)$ instead of $iv_{\text{\tiny{F}}}\tau$ in denominator of (\ref{logProp-3}) (see Ref.~\onlinecite{Lidsky}). To reproduce it, we would need more complex form of $G(k, \omega_m)$ than the one we borrowed from Ref.~\onlinecite{AIM}.


\begin{widetext}
\section{Two-loop contribution to the self-energy at a QCP.}
\label{two-loop}
In this Appendix, we show the details of the calculations of the
two-loop and three-loop self-energy diagrams.

\subsection{Two-loop diagram}

The diagram is presented in Fig.~\ref{fig:se2}.
In analytic form
\begin{eqnarray}
\label{1sigma2}
\Sigma_2(k,\omega_m)&=&
 -
g^4\int \frac{d \omega
_1 \;d{\bf q}_1}{(2\pi)^3}\int \frac{d\omega_2\;d{\bf
q}_2}{(2\pi)^3}\;\chi({\bf q}_1,\omega_1)\chi({\bf
q}_2,\omega_2)
G({\bf k}+{\bf q}_1,\omega_m-\omega_1)\nonumber\\
& \times &
G({\bf k}+{\bf q}_2,\omega_m-\omega_2)
G({\bf k}+{\bf
q}_1+{\bf q}_2,\omega_m-\omega_1-\omega_2)\\
 &= & \frac{\bar{g}^2}{(2\pi)^6} \int d\omega_1 d {\bf q}_1 \int d\omega_2 d {\bf q}_2
 \frac{|q_1|}{\gamma |\omega_1|+ |q_1|^3}
 \frac{|q_2|}{\gamma |\omega_2|+|q_2|^3} \nonumber \\
& \times &\left[\frac{1}{\tilde{\Sigma}(\omega_m-\omega_1)-v_{\text{\tiny{F}}}
(k_x+q_{1x})-N(k_y+q_{1y})^2/2m}\right]
\left[\frac{1}{\tilde{\Sigma}(\omega_m-\omega_2)-v_{\text{\tiny{F}}} (k_x+q_{2x})-N(k_y+q_{2y})^2/2m}\right] \nonumber \\
& \times &\left[\frac{1}{
\tilde{\Sigma}(\omega_m-\omega_1-\omega_2)-v_{\text{\tiny{F}}} (k_x+q_{1x}+
q_{2x})-N(k_y+q_{1y}+q_{2y})^2/2m}\right] .\nonumber
\end{eqnarray}
Integrating over
$q_{1x}$ and $q_{2x}$ and rescaling frequencies $\omega_1$,
$\omega_2$ by $\omega_m$ and momenta $q_{1y},\; q_{2y}$ by $(\gamma
\omega_{1,2})^{1/3}$, we obtain
\begin{eqnarray}
 \label{1sigma2-1}
\Sigma_2(k,\omega_m) &=& -\left(\frac{\sqrt{3}}{2\pi}\right)^2\omega_m^{2/3}
\omega_0^{1/3}\int_{-\infty}^{\infty} \frac{d \bar{q}_1 d
\bar{q}_2 |\bar{q}_1
\bar{q}_2|}{(1+|\bar{q}_1|^3)(1+|\bar{q}_2|^3)} \int_0^1\frac{d
\Omega_1}{\Omega_1^{1/3}}\int_{1-\Omega_1}^1\frac{d
\Omega_2}{\Omega_2^{1/3}}
\\
&\times& \frac{1}{2 \sqrt{3} N \bar{q}_1
\bar{q}_2(\Omega_1\Omega_2)^{1/3}+\left(i-\frac{\epsilon_k}{\omega_m^{2/3}
\omega_0^{1/3}}\right) +i
\left[(1-\Omega_1)^{2/3}+(1-\Omega_2)^{2/3}+(\Omega_1+\Omega_2-1)^{2/3}-1
\right] },\nonumber
\end{eqnarray}
where $\Omega_{1,2}=\frac{\omega_{1,2}}{\omega_m}$ and
$\bar{q}_{1,2}=\frac{q_{(1,2)y}}{(\gamma \omega_{1,2})^{1/3}}$.
\end{widetext}

\subsubsection{Large N expansion}

The two-loop self-energy to leading order in $1/N$ is obtained by
 expanding the denominator in (\ref{1sigma2-1}) to order $1/N^2$. The two momentum integrals are logarithmical with the lower limit set by $1/N$.
We have
\begin{eqnarray}
\label{quadratic1}
\Sigma_2 (k,\omega_m)=-\left(\frac{\ln^2N}{16\pi^2N^2}\right)\omega_m^{2/3}\omega_0^{1/3}
\int_0^1\frac{d\Omega_1}{\Omega_1}
\int_{1-\Omega_1}^{1}\frac{d\Omega_2}{\Omega_2}\nonumber\\
\times\Biggl\{\frac{\epsilon_k}{\omega_0^{1/3}\omega_m^{2/3}}-
i\Bigl[(1-\Omega_1)^{2/3}+(1-\Omega_2)^{2/3}\qquad\qquad\;\;\nonumber\\
+(\Omega_1+\Omega_2-1)^{2/3}
\Bigr]\Biggr\}. \qquad
\end{eqnarray}
Integrating over $\Omega_1$ and $\Omega_2$ in Eq.~(\ref{quadratic1})
 we obtain  Eq.~(\ref{quadratic2}) of the main text.

\subsubsection{Expansion near the mass-shell}
\label{Nf2}

Eq.~(\ref{quadratic1}) is valid when $\ln N$ is large and the result,
Eq.~(\ref{quadratic2}), contains only the term linear in $\Delta$.
 For small $N$, the momentum integration has to be done more accurately.
Below we expand $\Sigma_2$ near the mass shell and show that this expansion contains  a non-analytic $\Delta^{5/2}$ term.  This non-analytic term is
comes from low-energy fermions and is enhanced by vertex corrections.

We assume and then verify that the non-analytic term comes from
the regions $\Omega_1 \ll 1$,
$1-\Omega_2 \ll 1$ and $\Omega_1 \ll 1$,
$1-\Omega_2 \ll 1$. The contributions from these two regions are equal and we only focus on the contribution from  $\Omega_1 \ll 1$,
$1-\Omega_2 \ll 1$. We introduce $\Omega^{\prime}_2=1-\Omega_2$ and
 make use of the identity
\begin{eqnarray}
\label{doubleint}
&&\int_{-\infty}^{\infty}\frac{dq_1|q_1|}{1+|q_1|^3}\int_{-\infty}^{\infty}\frac{dq_2|q_2|}{1+|q_2|^3}
\frac{1}{A+iBq_1q_2}\\
&&=\frac{A}{54(A^6+B^6)}\Biggl[\pi^2\left(32A^4-32A^2B^2+17B^4\right)\nonumber\\
&&\qquad+108B\ln\left({B}/{A}\right)\biggl\{\pi A^3+B^3\ln(B/A)\biggr\}\Biggr],\nonumber
\end{eqnarray}
where in our case
\begin{eqnarray}
\label{purposes}
B = 2\sqrt{3}N(\Omega_1)^{1/3},\;A =\Delta+\left(\Omega_2^{\prime}\right)^{2/3}+\left(\Omega_1-\Omega_2^{\prime}\right)^{2/3}.\nonumber\\
\end{eqnarray}
The source for non-analyticity is the term with
 $\ln^2\left(B/{A}\right)$, and keeping only this term we obtain
\begin{eqnarray}
\label{univ2_1}
&&\!\!\!\!\!\!\!\Sigma_2(k,\omega_m) = 2i \omega_m^{2/3}
\omega_0^{1/3}\left(\frac{\sqrt{3}}{2\pi}\right)^2\int_0^1\!\frac{d\Omega_1}{\Omega_1^{1/3}}
\int_0^{\Omega_1}d\Omega_2^{\prime}\nonumber\\
&&\!\!\!\!\!\!\!
\times\frac{2\left[\Delta+\left(\Omega_2^{\prime}\right)^{2/3}+\left(\Omega_1-\Omega_2^{\prime}\right)^{2/3}
\right]\left(2\sqrt{3}N\Omega_1^{1/3}\right)^4}
{\left[\Delta+\left(\Omega_2^{\prime}\right)^{2/3}+\left(\Omega_1-\Omega_2^{\prime}\right)^{2/3}\right]^6+\left(2\sqrt{3}N\Omega_1^{1/3}\right)^6}\nonumber\\
&&\times\ln^2\left[\frac{N\Omega_1^{1/3}}{\Delta+\left(\Omega_2^{\prime}\right)^{2/3}+\left(\Omega_1-\Omega_2^{\prime}\right)^{2/3}}\right].
\end{eqnarray}
Introducing new integration variables, $x=\Omega_1/\Delta^{3/2}$, $y=\Omega_2^{\prime}/\Delta^{3/2}$,  we re-write Eq.~(\ref{univ2}) as
\begin{widetext}
\begin{eqnarray}
\label{univ2_2}
&&\!\!\!\!\Sigma_2(k,\omega_m) = i \omega_m^{2/3}
\omega_0^{1/3}\frac{\Delta^{5/2}}{4 \pi^2 N^2}
\int_0^{1/\Delta^{3/2}}\!\frac{dx}{x}\int_0^{x} dy
\left[1+y^{2/3}+(x-y)^{2/3}\right] \Biggl[\ln\left(\frac{N}{\sqrt{\Delta}}\right)+\ln\left(\frac{x^{1/3}}{1+y^{2/3}+(x-y)^{2/3}}\right)\Biggr]^2.\nonumber\\
\end{eqnarray}
\end{widetext}
This expression contains contributions confined to the upper limit of the integration over $x$, but also contains universal contributions which come from $x, y = O(1)$ and is independent on the upper limit of integration. One can easily verify that the largest contribution of this kind
 comes from the cross-product of the two logarithmic factors in the last
bracket. Evaluating the integrals explicitly we obtain
 Eq.~(\ref{univ2}) of the main text.

\subsection{Three-loop diagram}
\label{three-loop}

We verified that the most singular contributions to fermionic
self-energy at the three-loop level come from the two diagrams
presented in Fig.~\ref{fig:se2}b. Compared to the two-loop
 diagram, these two diagrams describe corrections to the two spin-boson vertices which involve a boson with a frequency $\omega_2 \approx \omega$.
 Both diagrams give equal contributions, and we consider only the one from Fig.~\ref{fig:se2}b.
The analytical expression for this diagram is
\begin{widetext}
\begin{eqnarray}
\label{1sigma3} \Sigma_3(k,\omega_m)&=& g^6\int \frac{d \omega
_1 \;d{\bf q}_1}{(2\pi)^3}\int \frac{d\omega_2\;d{\bf
q}_2}{(2\pi)^3}\int \frac{d \omega _3 \;d{\bf
q}_3}{(2\pi)^3}\;\chi({\bf q}_1,\omega_1)\chi({\bf
q}_2,\omega_2)\chi({\bf q}_3,\omega_3)G({\bf k}+{\bf
q}_1,\omega_m-\omega_1)\nonumber\\
& \times & G({\bf k}+{\bf q}_1+{\bf
q}_2,\omega_m-\omega_1-\omega_2)G({\bf k}+{\bf
q}_2,\omega_m-\omega_2)G({\bf k}+{\bf q}_2+{\bf
q}_3,\omega_m-\omega_2-\omega_3)G({\bf k}+{\bf
q}_3,\omega_m-\omega_3)\\
 &= & \frac{\bar{g}^3}{(2\pi)^6} \int d\omega_1 d {\bf q}_1 \int d\omega_2 d {\bf q}_2\int d\omega_3 d {\bf q}_3
 \frac{|q_1|}{\gamma |\omega_1|+ |q_1|^3}
 \frac{|q_2|}{\gamma |\omega_2|+|q_2|^3}
 \frac{|q_3|}{\gamma |\omega_3|+ |q_3|^3}
 \nonumber \\
& \times &\left[\frac{1}{\tilde{\Sigma}(\omega_m-\omega_1)-v_{\text{\tiny{F}}}
(k_x+q_{1x})-N(k_y+q_{1y})^2/2m}\right]
\left[\frac{1}{\tilde{\Sigma}(\omega_m-\omega_2)-v_{\text{\tiny{F}}} (k_x+q_{2x})-N(k_y+q_{2y})^2/2m}\right] \nonumber \\
&\times &\left[\frac{1}{\tilde{\Sigma}(\omega_m-\omega_3)-v_{\text{\tiny{F}}}
(k_x+q_{3x})-N(k_y+q_{3y})^2/2m}\right]\nonumber\\
& \times &\left[\frac{1}{
\tilde{\Sigma}(\omega_m-\omega_1-\omega_2)-v_{\text{\tiny{F}}} (k_x+q_{1x}+
q_{2x})-N(k_y+q_{1y}+q_{2y})^2/2m}\right]\nonumber\\
& \times &\left[\frac{1}{
\tilde{\Sigma}(\omega_m-\omega_2-\omega_3)-v_{\text{\tiny{F}}} (k_x+q_{2x}+
q_{3x})-N(k_y+q_{2y}+q_{3y})^2/2m}\right].
\end{eqnarray}
Performing the integration over variables $q_{1x}$, $q_{2x}$, and
$q_{3x}$ we obtain
\begin{eqnarray}
 \label{1sigma3-1}
\Sigma_3(k,\omega_m) &=&
-\left(\frac{\sqrt{3}}{2\pi}\right)^3\omega_m^{2/3}
\omega_0^{1/3}\int_{-\infty}^{\infty} \frac{d \bar{q}_1 d
\bar{q}_2 d \bar{q}_3|\bar{q}_1
\bar{q}_2\bar{q}_3|}{(1+|\bar{q}_1|^3)(1+|\bar{q}_2|^3)(1+|\bar{q}_3|^3)}
\int_0^1\frac{d
\Omega_2}{\Omega_2^{1/3}}\int_{1-\Omega_2}^1\frac{d
\Omega_1}{\Omega_1^{1/3}} \int_{1-\Omega_2}^1\frac{d
\Omega_3}{\Omega_3^{1/3}}
\nonumber\\
&\times& \frac{1}{2 \sqrt{3} N \bar{q}_1
\bar{q}_2(\Omega_1\Omega_2)^{1/3}+i\Delta +i
\left[(1-\Omega_1)^{2/3}+(1-\Omega_2)^{2/3}+(\Omega_1+\Omega_2-1)^{2/3}-1
\right] }
\nonumber\\
&\times& \frac{1}{2 \sqrt{3} N \bar{q}_2
\bar{q}_3(\Omega_2\Omega_3)^{1/3}+i\Delta +i
\left[(1-\Omega_2)^{2/3}+(1-\Omega_3)^{2/3}+(\Omega_2+\Omega_3-1)^{2/3}-1
\right] } ,
\end{eqnarray}
where $\Omega_{1,2}=\frac{\omega_{1,2}}{\omega_m}$ and
$\bar{q}_{1,2}=\frac{q_{(1,2)y}}{(\gamma \omega_{1,2})^{1/3}}$.
Compared to the two-loop diagram given by
 Eq.~(\ref{1sigma2-2}) this one contains an additional integral
 over $\Omega_3$ and additional denominator.

We are interested in the corrections to the universal, non-analytic term in
the self-energy. Accordingly, we introduce, as before,
$\Omega_2^{\prime}=1 - \Omega_2$ and set
$\Omega^\prime_2, \Omega_1 \ll 1$.  We then obtain
\begin{eqnarray}
 \label{1sigma3-2}
\Sigma_3(k,\omega_m)\!\! &=&\!\!
 \;i\omega_m^{2/3}
\omega_0^{1/3}\left(\frac{\sqrt{3}}{2\pi}\right)^3
\int_{-\infty}^{\infty} \frac{d \bar{q}_1 d
\bar{q}_2 d \bar{q}_3|\bar{q}_1
\bar{q}_2\bar{q}_3|}{(1+|\bar{q}_1|^3)(1+|\bar{q}_2|^3)(1+|\bar{q}_3|^3)}
\int_{0}^1 d \Omega_2^{\prime} \int_{\Omega_2^{\prime}}^1 \frac{d \Omega_1}{\Omega_1^{1/3}}
\int_{\Omega_2^{\prime}}^{1}\frac{d \Omega_3}{\Omega_3^{1/3}}\\
\!\!&\times&\!\!
\frac{1}{-\left(2 \sqrt{3}\;i\right) N \bar{q}_1
\bar{q}_2\Omega_1^{1/3}+\Delta +
\left[\Omega_2^{\prime\; 2/3}+(\Omega_1-\Omega_2^{\prime})^{2/3}
\right] }\;
\frac{1}{-\left(2 \sqrt{3}\;i\right) N \bar{q}_2
\bar{q}_3\Omega_3^{1/3}+\Delta +
\left[\Omega_2^{\prime\; 2/3}+ (\Omega_3-\Omega_2^{\prime})^{2/3}
\right] }.\nonumber
\end{eqnarray}

\subsubsection{$N=0$}

Integrating over dimensionless momenta $\bar{q}_1$
and $\bar{q}_2$ in Eq.~(\ref{1sigma3-2}) and setting  $N=0$ we obtain
\begin{eqnarray}
 \label{zeroN3}
\Sigma_3(k,\omega_m)&=&\frac{4}{9}\;i\;\omega_m^{2/3}
\omega_0^{1/3}
\int_{0}^1\frac{d
\Omega_1}{\Omega_1^{1/3}}
\int_0^{\Omega_1} d\Omega_2^{\prime}\frac{1}{
\Delta +
\left[\Omega_2^{\prime\; 2/3}+(\Omega_1-\Omega_2^{\prime})^{2/3}
\right]} \nonumber \\
&& \times \frac{2}{3}
\int_{\Omega^{\prime}_2}^1\frac{d\Omega_3}{\Omega_3^{1/3}}\frac{1}{\Delta +
\left[\Omega_2^{\prime\; 2/3}+ (\Omega_3-\Omega_2^{\prime})^{2/3}
\right] }.
\nonumber\\
\end{eqnarray}
\end{widetext}
The first line is the expression for one of the two two-loop self-energy diagram, the second line is the extra piece which represents the vertex correction.
We see that the integral over $\Omega_3$ is logarithmical. Using the fact that
 typical $\Omega'_2 \sim \Delta^{3/2}$, we obtain, with logarithmical accuracy,
\be
\label{zeroNsigma3}
\Sigma_3(k,\omega_m)= \Sigma_2(k,\omega_m)\ln{\frac{1}{\Delta}}.
\ee
This is the result that we cited in the main text.

\subsubsection{Finite $N$}

At a finite $N$ and $\Delta \ll N^2$, the logarithmic divergence of the
 integral over $\Omega_3$ in (\ref{zeroN3}) is cut by $N^2$ instead of $\Delta$, and we have
\begin{eqnarray}
\label{zeroNsigma3_1}
\Sigma_3(k,\omega_m)&=& \Sigma_2(k,\omega_m)\ln{\frac{1}{N^2}}.
\end{eqnarray}


\section{Fermions at the half-filled Landau level}
\label{CS}
In this Appendix we give details of the calculations for a model of
 fermions interacting with a bosonic field whose static propagator scales as
$1/|q|$.

\subsection{Real-space propagator for $N\gg 1$}
\label{z=2}

The derivation  of the form of $G(r,t)$ is analogous to the derivation
 of Eq.~(\ref{NMN}). Using the self-energy $\Sigma (\omega_m)= i \lambda \omega_m \ln\left(\epsilon_0/\omega_m\right)$ and converting it to real frequencies, we obtain, with logarithmical accuracy,
\begin{eqnarray}
\label{rt-rep}
G({\bf r},t)=\int_{-\infty}^{\infty}\frac{d\omega}{2\pi}\int\frac{d{\bf k}}{(2\pi)^2}\;\frac{\exp\left(i{\bf k} {\bf r}-i\omega t\right)}{\lambda
\omega\ln\left(\epsilon_0/\omega\right)-\epsilon_k}.\nonumber\\
\end{eqnarray}
Introducing new variables
 $x=\lambda\omega\ln\left(\epsilon_0/\omega\right)/\epsilon_k$,
$y=\epsilon_kr/v_{\text{\tiny F}}$ and using the fact that
\begin{eqnarray}
\label{intid}
\int_0^{\infty}dy\;\cos(a+y)e^{-by}=\frac{b\cos a-\sin a}{1+b^2},\;\;\;\;b>0,
\end{eqnarray}
we arrive at Eq.~(\ref{compare}).

\subsection{Two-loop diagram}

The analytical expression for $\Sigma_2(q,\omega_m)$ has the same structure as
 Eq.~(\ref{1sigma2}) and after the integration over $x-$component of momenta reduces to
\begin{widetext}
\begin{eqnarray}
 \label{1sigma2-11}
\Sigma_2(k,\omega_m) &=& - \lambda^2
\int_0^{\omega_m}d\omega_1\int_{\omega_m-\omega_1}^{\omega_m}d\omega_2
\int\frac{d {q}_1 d{q}_2 |{q}_1 {q}_2|}
{(\tilde\gamma \omega_1+|{q}_1|^2)(\tilde\gamma \omega_2+|{q}_2|^2)}
\nonumber\\
&\times& \frac{1}{ \left(Nq_1 q_2/m\right)-\epsilon_k
+\left[\tilde{\Sigma}(\omega_m-\omega_1)+\tilde{\Sigma}(\omega_m-\omega_2)+
\tilde{\Sigma}(\omega_1+\omega_2-\omega_m) \right] },
\end{eqnarray}
where $\tilde\gamma=\gamma/(mu)=4\pi\lambda m$ and
 $q_{1,2}$ are $y$ components of running momenta bounded by
$|q_{1,2}|\leq k_{\text{\tiny{F}}}$.

\subsubsection{Large N expansion}

As in the previous case, we expand the  denominator in Eq.~(\ref{1sigma2-11}) in powers of $(1/N)\ll 1$ and cut the logarithms by $1/N$.  To leading order in $1/N$, we have
\begin{eqnarray}
\label{quadratic2_2}
\Sigma_2 (k,\omega_m)=\left[\frac{\ln^2N}{4\pi^2N^2}\right]\int_0^{\omega_m}\frac{d\omega_1}{\omega_1}\int_{\omega_m-\omega_1}^{\omega_m}
\frac{d\omega_2}{\omega_2}
\Biggl\{-\epsilon_k
&+&i\lambda\Biggl[(\omega_m-\omega_1) \ln\left(\frac{\epsilon_0}{\omega_m-\omega_1}\right)
+(\omega_m-\omega_2) \ln\left(\frac{\epsilon_0}{\omega_m-\omega_2}\right)\nonumber\\
&+&(\omega_1+\omega_2-\omega_m) \ln\left(\frac{\epsilon_0}{\omega_1+\omega_2-\omega_m}\right)\Biggr]\Biggr\}.
\end{eqnarray}
Using
\begin{eqnarray}
\label{epsilonk}
\int_0^{\omega_m}\frac{d\omega_1}{\omega_1}\int_{\omega_m-\omega_1}^{\omega_m}
\frac{d\omega_2}{\omega_2}=\frac{\pi^2}{6},
\end{eqnarray}
and
\begin{eqnarray}
\label{omegaLn}
\int_0^{\omega_m}\frac{d\omega_1}{\omega_1}\int_{\omega_m-\omega_1}^{\omega_m}
\frac{d\omega_2}{\omega_2}\Biggl[(\omega_m-\omega_1) \ln\left(\frac{\epsilon_0}{\omega_m-\omega_1}\right)
+(\omega_m-\omega_2) \ln\left(\frac{\epsilon_0}{\omega_m-\omega_2}\right)\nonumber\\
+(\omega_1+\omega_2-\omega_m) \ln\left(\frac{\epsilon_0}{\omega_1+\omega_2-\omega_m}\right)\Biggr]
=\frac{\pi^2}{6}\;\omega_m\ln(\epsilon_0/\omega_m),
\end{eqnarray}
and substituting Eqs.~(\ref{epsilonk}) and (\ref{omegaLn}) into Eq.~(\ref{quadratic2}) we obtain Eq.~(\ref{MargSigma2}) of the main text.

\subsubsection{Universal term in the expansion near the mass-shell}
As in Appendix \ref{two-loop}  we introduce
new dimensionless variables $\bar{q}_1 = q_1/\sqrt{\tilde{\gamma}
\omega_1}$ and $\bar{q}_2 = q_2/\sqrt{\tilde{\gamma}
\omega_2}$. The integration over $\bar{q}_1$ and
$\bar{q}_2$ in Eq.~(\ref{1sigma2-11}) is now confined to
$|\bar{q}_{1,2}|\leq \left(\epsilon_0/\omega_{1,2}\right)^{1/2}$,
and Eq.~(\ref{1sigma2-11}) becomes, after proper rescaling
\begin{eqnarray}
\label{sigma2NCS}
&&\Sigma_2(k,\omega_m)=i\lambda\;\omega_m\int_0^{1}d\Omega_1\int_{1-\Omega_1}^{1}
d\Omega_2\int\frac{d\bar{q}_1 |\bar{q}_1|}{1+\bar{q}_1^2}
\int\frac{d\bar{q}_2 |\bar{q}_2|}{1+\bar{q}_2^2}\times\\
&&\frac{1}{-iN\alpha\sqrt{\Omega_1\Omega_2}\bar{q}_1\bar{q}_2+\tilde{\Delta}\ln\left(\epsilon_0/\omega_m\right)+
\left\{(1-\Omega_1)\ln\left[\frac{1}{1-\Omega_1}\right]+(1-\Omega_2)\ln\left[\frac{1}{1-\Omega_2}\right]
+(\Omega_1+\Omega_2-1)\ln\left[\frac{1}{\Omega_1+\Omega_2-1}\right]\right\}},
\nonumber
\end{eqnarray}
where $\Omega_{1,2}=\left(\omega_{1,2}/\omega_m\right)$, $\alpha=(1/2\pi)(v_F/u)$ and ${\tilde \Delta}$ is given by
Eq.~(\ref{Dtilde}).

As before, we expand $\Sigma_2$ near the mass-shell in
powers of $\tilde{\Delta}$ and search for a universal, non-analytical contributions from $\Omega_1\sim 0$,
$\Omega_2\sim 1$ and vice versa. Restricting with the contribution from the first region and introducing  $\Omega^{\prime}_2=1-\Omega_2 \ll 1$, we obtain
\begin{eqnarray}
\label{N2CS}
\Sigma_2(k,\omega_m)&=&i\lambda\;\omega_m\int_{0}^{1}
d\Omega_2^{\prime}\int_{\Omega_2^{\prime}}^{1}d\Omega_1\int\frac{d\bar{q}_1 |\bar{q}_1|}{1+\bar{q}_1^2}
\int\frac{d\bar{q}_2 |\bar{q}_2|}{1+\bar{q}_2^2}
\nonumber\\
&\times&
\;\frac{1}{-iN\alpha\sqrt{\Omega_1}\bar{q}_1\bar{q}_2+\tilde{\Delta}\ln\left(\epsilon_0/\omega_m\right)+
\Omega_2^{\prime}\ln\left(\frac{1}{\Omega_2^{\prime}}\right)
+\left(\Omega_1-\Omega_2^{\prime}\right)\ln\left(\frac{1}{\Omega_1-\Omega_2^{\prime}}\right)}.\;\;\nonumber\\
\end{eqnarray}
\end{widetext}
The integrals over $\bar{q}_1$ and
$\bar{q}_2$ are confined to
$|\bar{q}_{1}|\leq Q_1=\left(\epsilon_0/\omega_m\right)^{1/2}(1/\Omega_1)^{1/2}$ and $|\bar{q}_{2}|\leq Q_2=\left(\epsilon_0/\omega_m\right)^{1/2}$, respectively,
the upper limits of the integrals over $\Omega_1$ and $\Omega'_2$ should not matter in this approximation.

A simple experimentation shows that the universal, non-analytic contribution to $\Sigma_2$ appears already at first order in ${\tilde \Delta}$. Expanding
(\ref{N2CS}) in ${\tilde \Delta}$ and integrating over $\bar{q}_1$ and $\bar{q}_2$ we obtain with  the logarithmic accuracy
\vspace{0.5cm}
\begin{eqnarray}
\label{NfCS2}
\Sigma_2(k,\omega_m)&=& -i\lambda\;\omega_m\; \tilde{\Delta}\ln^2\left(\epsilon_0/\omega_m\right)\int_{\Omega_0}^{1}
d\Omega_1\int_{\Omega_0}^{\Omega_1}d\Omega'_2\times\nonumber\\
&&\frac{\ln\left(1/\Omega_2^{\prime}\right)}{
\left[
\Omega_2^{\prime}\ln\left(\frac{1}{\Omega_2^{\prime}}\right)
+\left(\Omega_1-\Omega_2^{\prime}\right)\ln\left(\frac{1}{\Omega_1-\Omega_2^{\prime}}\right)\right]^2}\nonumber\\
&=& -i\lambda\;\omega_m\; \tilde{\Delta}\ln^2\left(\epsilon_0/\omega_m\right) \ln{\ln{\frac{1}{\Omega_0}}}.
\end{eqnarray}
The lower cutoff $\Omega_0$ is defined by $\Omega_0
\ln (1/\Omega_0) \sim {\tilde \Delta} \ln{\epsilon_0/\omega_m}$ at $N=0$
and by $\Omega_0 \ln^2 (1/\Omega_0) \sim N^2 \ln^4 N$ at  $N^2 \gg {\tilde \Delta}$ (up to extra logarithms).
This leads to Eqs.~(\ref{CS3}) and (\ref{CSS2}) in the main text.


\end{document}